%% file: main.tex
\documentclass[twocolumn]{aastex631}

\begin{document}

\title{The SDSS-V Local Volume Mapper (LVM): Technical Overview}

\input{authors}




\begin{abstract}

The Sloan Digital Sky Survey V (SDSS-V) Local Volume Mapper (LVM) is an ultra-wide integral-field spectroscopic survey of the Milky Way, the Magellanic Clouds, and a sample of galaxies in the local volume. The LVM spatially resolved spectroscopy covers the bulk of the optical range (3600--9800~\AA) at medium spectral-resolution ($R=2200-5500$), and maps the disks of these galaxies at high physical spatial resolution ($<1$~pc in the MW and $\sim10$~pc in the Magellanic Clouds). Observations are carried out with the LVM Instrument (LVM-I), a custom-built robotic telescope, instrument, and facility located at Las Campanas Observatory (LCO) in Chile. In this article we present a technical overview of the LVM project. We describe the hardware, the survey observing strategy, and the details of how science and calibration observations are carried out. We provide a high-level overview of the survey data flow, including the reduction and calibration of the LVM spectra, and the production of the main data products that are released by the project. We also present an assessment of the LVM data quality and instrumental stability based on science observations. The main goal of this article is to serve as a primer for LVM users, providing the technical information required for the adequate analysis, modeling, and interpretation of the LVM data.

\end{abstract}



\input{1-intro}

\input{2-instrument}

\input{3-footprint}

\input{4-observing}

\input{5-data}

\input{6-products}

\input{7-data_quality}

\input{8-conclusions}

\begin{acknowledgements}

Funding for the Sloan Digital Sky Survey V has been provided by the Alfred P. Sloan Foundation, the Heising-Simons Foundation, the National Science Foundation, and the Participating Institutions. SDSS acknowledges support and resources from the Center for High-Performance Computing at the University of Utah. SDSS telescopes are located at Apache Point Observatory, funded by the Astrophysical Research Consortium and operated by New Mexico State University, and at Las Campanas Observatory, operated by the Carnegie Institution for Science. The SDSS web site is \url{www.sdss.org}.

SDSS is managed by the Astrophysical Research Consortium for the Participating Institutions of the SDSS Collaboration, including Caltech, The Carnegie Institution for Science, Chilean National Time Allocation Committee (CNTAC) ratified researchers, The Flatiron Institute, the Gotham Participation Group, Harvard University, Heidelberg University, The Johns Hopkins University, L’Ecole polytechnique f\'{e}d\'{e}rale de Lausanne (EPFL), Leibniz-Institut f\"{u}r Astrophysik Potsdam (AIP), Max-Planck-Institut f\"{u}r Astronomie (MPIA Heidelberg), Max-Planck-Institut f\"{u}r Extraterrestrische Physik (MPE), Nanjing University, National Astronomical Observatories of China (NAOC), New Mexico State University, The Ohio State University, Pennsylvania State University, Smithsonian Astrophysical Observatory, Space Telescope Science Institute (STScI), the Stellar Astrophysics Participation Group, Universidad Nacional Aut\'{o}noma de M\'{e}xico, University of Arizona, University of Colorado Boulder, University of Illinois at Urbana-Champaign, University of Toronto, University of Utah, University of Virginia, Yale University, and Yunnan University.

G.A.B. and E.J.J acknowledge the support from the ANID Basal project FB210003. 

E.J.J. acknowledges support from FONDECYT Iniciaci\'on en investigaci\'on 2020 Project 11200263 and the Millenium Nucleus ERIS NCN2021\_017, Centros ANID Iniciativa Milenio.

K.K., E.E, M.C.iS., H.G., J.L., S.A.P. and N.S. gratefully acknowledge funding from the Deutsche Forschungsgemeinschaft (DFG, German Research Foundation) in the form of an Emmy Noether Research Group (grant number KR4598/2-1, PI Kreckel) and the European Research Council’s starting grant ERC StG-101077573 (“ISM-METALS"). 

O.E. acknowledges funding from the Deutsche Forschungsgemeinschaft (DFG, German Research Foundation) -- project-ID 541068876.

This work has made use of the Python package GaiaXPy, developed and maintained by members of the Gaia Data Processing and Analysis Consortium (DPAC) and in  particular, Coordination Unit 5 (CU5), and the Data Processing Centre located at the Institute of Astronomy, Cambridge, UK (DPCI).

\end{acknowledgements}

\bibliography{lvm}{}
\bibliographystyle{aasjournal}



\end{document}

%% file: authors.tex
\newcommand{\Carnegie}{\affiliation{The Observatories of the Carnegie Institution for Science, 813 Santa Barbara Street, Pasadena, CA 91101, USA}}

\newcommand{\UChile}{\affiliation{Departamento de Astronom\'{i}a, Universidad de Chile, Camino del Observatorio 1515, Las Condes, Santiago, Chile}}

\newcommand{\MDO}{\affiliation{McDonald Observatory, The University of Texas at Austin, 1 University Station, Austin, TX 78712-0259, USA}
}
\newcommand{\UNAMDF}{\affiliation{Instituto de Astronom\'ia, Universidad Nacional Aut\'onoma de M\'exico, A.P.\ 70-264, 04510 M\'exico D.\ F., M\'exico}}

\newcommand{\UNAMEnsenada}{\affiliation{Instituto de Astronom\'ia, Universidad Nacional Aut\'onoma de M\'exico, A.P.\ 106, Ensenada 22800, BC, M\'exico}}

\newcommand{\Heidelberg}{\affiliation{Astronomisches Rechen-Institut, Zentrum f\"ur Astronomie der Universit\"at Heidelberg, M\"onchhofstr.\ 12-14, D-69120 Heidelberg, Germany}}

\newcommand{\Illinois}{\affiliation{Department of Astronomy, University of Illinois, Urbana, IL 61801, USA}}

\newcommand{\ITA}{\affiliation{Institut fur theoretische Astrophysik, Zentrum f\"ur Astronomie der Universit\"at Heidelberg, Albert-Ueberle-Str.\ 2, D-69120 Heidelberg, Germany}}

\newcommand{\SWRI}{\affiliation{Southwest Research Institute, 6220 Culebra Road, San Antonio, TX 78238  }}

\newcommand{\MPIA}{\affiliation{Max Planck Institute for Astronomy, K\"onigstuhl 17, D-69117 Heidelberg, Germany}
}

\newcommand{\UDP}{\affiliation{Instituto de Estudios Astrof\'isicos, Facultad de Ingenier\'ia y Ciencias, Universidad Diego Portales, Av.\ Ej\'ercito Libertador 441, Santiago, Chile.}}

\newcommand{\STScI}{\affiliation{Space Telescope Science Institute, 3700 San Martin Drive, Baltimore, MD 21218, USA}
}

\newcommand{\NYUAbuDhabi}{\affiliation{New York University Abu Dhabi, PO Box 129188, Abu Dhabi, UAE}}

\newcommand{\CASS}{\affiliation{Center for Astrophysics and Space Science (CASS), New York University Abu Dhabi, PO Box 129188, Abu Dhabi, UAE}}

\newcommand{\Sternberg}{\affiliation{Sternberg Astronomical Institute, Lomonosov Moscow State University, Universitetskij pr., 13,  Moscow, 119234, Russia}}

\newcommand{\UW}{\affiliation{Department of Astronomy, University of Washington, Box 351580, Seattle, WA 98195, USA}
}
\newcommand{\APO}{\affiliation{Apache Point Observatory and New Mexico State University, P.O.\ Box 59, Sunspot, NM 88349-0059, USA}
}

\newcommand{\ESOChile}{\affiliation{European Southern Observatory, Alonso de C\'ordova 3107, Casilla 19, Santiago 19001, Chile}
}

\newcommand{\UNAB}{\affiliation{Universidad Andres Bello, Facultad de Ciencias Exactas, Departamento de Ciencias F\'isicas, Instituto de Astrof\'isica, Av. Fern\'andez Concha 700, Santiago, Chile}
}

\newcommand{\TCU}{\affiliation{Department of Physics \& Astronomy, Texas Christian University, Fort Worth, TX 76129, USA}
}

\newcommand{\ESO}{\affiliation{European Southern Observatory, Karl-Schwarzschild-Straße 2, 85748 Garching, Germany}
}

\newcommand{\UDA}{\affiliation{Instituto de Astronom\'ia y Ciencias Planetarias de Atacama, Universidad de Atacama, Copayapu 485, Copiap\'o, Chile}
}

\newcommand{\AIP}{\affiliation{Leibniz-Institut f\"ur Astrophysik Potsdam, An der Sternwarte 16, D-14482 Potsdam, Germany}
}

\newcommand{\UCN}{\affiliation{Instituto de Astronom\'ia, Universidad Cat\'olica del Norte, Av.\ Angamos 0610, Antofagasta, Chile}
}

\newcommand{\CASSACA}{\affiliation{Chinese Academy of Sciences South America Center for Astronomy, National Astronomical Observatories, CAS, Beijing 100101, China}
}

\newcommand{\Monash}{\affiliation{School of Physics \& Astronomy, Monash University, Wellington Road, Clayton, Victoria 3800, Australia}
}

\newcommand{\ColoradoCollege}{\affiliation{Department of Physics, Colorado College, 14 East Cache la Poudre St., Colorado Springs, CO, 80903, USA}
}

\newcommand{\UCBoulder}{\affiliation{Department of Astrophysical and Planetary Sciences, University of Colorado, 389 UCB, Boulder, CO 80309-0389, USA}
}

\newcommand{\UDEC}{\affiliation{Departamento de Astronom\'ia, Universidad de Concepci\'on, Casilla 160-C, Concepci\'on, Chile}
}

\newcommand{\JodrellBank}{\affiliation{Jodrell Bank Centre for Astrophysics, Department of Physics and Astronomy, University of Manchester, Oxford Road, Manchester M13 9PL, UK}
}

\newcommand{\CITA}{\affiliation{Canadian Institute for Theoretical Astrophysics, University of Toronto, Toronto, ON M5S-98H, Canada}
}


\author[0000-0003-4218-3944]{Guillermo A. Blanc}
\Carnegie
\UChile
\email{gblancm@carnegiescience.edu}
\correspondingauthor{Guillermo Blanc}


\author[0000-0002-7339-3170]{Niv Drory}
\MDO

\author[0000-0001-6444-9307]{Sebasti\'an F.\ S\'anchez}
\UNAMEnsenada

\author[0000-0003-1905-2815]{Nicholas P.\ Konidaris}
\Carnegie

\author[0000-0001-6551-3091]{Kathryn Kreckel}
\Heidelberg

\author[0000-0002-0786-7307]{Tony Wong}
\Illinois

\author[0000-0003-3119-0815]{Eric W.\ Pellegrini}
\ITA


\author[0000-0001-8499-2892]{Cynthia S.\ Froning}
\SWRI

\author[0009-0009-8473-7205]{Tom Herbst}
\MPIA

\author[0000-0002-2368-6469]{Evelyn J.\ Johnston}
\UDP

\author[0000-0002-2262-8240]{Amy M.\ Jones}
\STScI

\author[0000-0002-6425-6879]{Ivan Yu. Katkov}
\NYUAbuDhabi
\CASS
\Sternberg

\author[0000-0002-8931-2398]{Alfredo Mej\'ia-Narv\'aez}
\UChile

\author[0009-0000-1165-4411]{Solange Ramirez}
\Carnegie

\author[0000-0003-2486-3858]{Jos\'e S\'anchez-Gallego}
\UW

\author[0009-0007-5938-3199]{Stefanie Wachter}
\Carnegie


\author[0000-0003-4996-9069]{Hans-Walter Rix}
\MPIA


\author[0000-0002-3601-133X]{Dmitry Bizyaev}
\APO

\author[0000-0002-4289-7923]{Brian Cherinka}
\STScI

\author[0000-0002-8549-4083]{Enrico Congiu}
\ESOChile

\author[0000-0002-2248-6107]{Maren Cosens}
\Carnegie

\author[0000-0003-4254-7111]{Bruno Dias}
\UNAB

\author[0009-0000-4049-5851]{John Donor}
\TCU

\author[0000-0002-4755-118X]{Oleg Egorov}
\Heidelberg

\author[0000-0003-2717-8784]{Evgeniya Egorova}
\Heidelberg

\author[0000-0002-5844-4443]{Maximilian H{\"a}berle}
\ESO

\author[0000-0002-9790-6313]{Hector Ibarra}
\UNAMDF

\author[0000-0002-4134-864X]{Knox S.\ Long}
\STScI

\author[0000-0002-6972-6411]{Jos\'e Eduardo M\'endez-Delgado}
\UNAMDF

\author[0009-0000-3962-103X]{Amrita Singh} 
\UChile


\author[0000-0003-2405-7258]{Jorge K.\ Barrera-Ballesteros}
\UNAMDF

\author[0000-0002-8518-6638]{Michelle Berg}
\TCU

\author[0009-0003-1585-9694]{Mar Canal i Saguer}
\Heidelberg

\author[0000-0002-6797-696X]{Maria-Rosa Cioni}
\AIP

\author[0009-0008-2218-3262]{Hugo Cort\'es-Mu\~noz}
\UChile

\author[0009-0008-7415-7015]{Rodrigo Cuellar}
\UChile

\author[0009-0002-5653-8814]{Florence de Almeida}
\UDP

\author[0009-0009-0081-4323]{Rodolfo de J. Zermeño}
\UNAMDF

\author[0000-0002-8297-6386]{Mariia Demianenko}
\MPIA

\author[0000-0002-8586-6721]{Pablo Garc\'ia}
\UCN
\CASSACA

\author[0000-0001-6708-1317]{Simon C. O. Glover}
\ITA

\author[0009-0001-4629-8098]{Hannah Greve}
\Heidelberg

\author[0000-0001-7641-5235]{Thomas Hilder}
\Monash

\author[0009-0004-5870-9562]{Kevin Hoy}
\UDP
\affiliation{Millennium Nucleus on Young Exoplanets and their Moons (YEMS)}

\author[0000-0002-4825-9367]{Jing Li}
\Heidelberg

\author[0000-0002-0560-3172]{Ralf S.\ Klessen}
\ITA
\affiliation{Universit\"{a}t Heidelberg, Interdisziplin\"{a}res Zentrum f\"{u}r Wissenschaftliches Rechnen, Im Neuenheimer Feld 205, D-69120 Heidelberg, Germany}

\author[0000-0002-7955-7359]{Dhanesh Krishnarao}
\ColoradoCollege

\author[0000-0002-5320-2568]{Nimisha Kumari}
\STScI

\author[0009-0003-9064-1089]{Sorya Lambert}
\UDP

\author[0000-0003-2496-1247]{Fu-Heng Liang}
\Heidelberg

\author[0009-0006-0286-0639]{Aurora Mata S\'anchez}
\UNAMDF

\author[0000-0002-7064-099X]{Dante Minniti}
\UNAB

\author[0009-0000-5681-3009]{Aparna Nair}
\UDA

\author[0009-0006-3186-5826]{Silvia Anastasia Popa}
\Heidelberg

\author[0000-0001-8858-1943]{Ravi Sankrit}
\STScI

\author[0000-0002-8883-6018]{Natascha Sattler}
\Heidelberg

\author[0009-0004-6638-338X]{Anna Lena Schaible}
\Heidelberg

\author[0000-0003-1479-3059]{Guy Stringfellow}
\UCBoulder

\author[0009-0008-2170-7845]{Luciano Vargas Herrera}
\UDEC

\author[0009-0008-1605-4771]{Monica Alejandra Villa Durango}
\UNAMEnsenada

\author[0000-0002-7365-5791]{Elizabeth Jayne Watkins}
\JodrellBank



\author[0000-0001-9852-1610]{Juna A.\  Kollmeier}
\Carnegie
\CITA

%% file: 1-intro.tex
\section{Introduction} \label{sec:intro}

The Local Volume Mapper \citep[LVM,][]{drory24} is one of the three programs that comprise the fifth realization of the Sloan Digital Sky Survey \citep[SDSS-V,][]{kollmeier26}. The main goal of the LVM project is to study the physical processes that act at the interface between the gas and stellar components of galaxies. These include: the formation of new stars out of the inter-stellar medium (ISM); the injection and propagation of energy and momentum that feeds back from these stars to the ISM, setting its kinematic, ionization, and thermodynamic state; and the chemical enrichment of interstellar gas from the dispersal of heavy elements produced in stars. 

To achieve this goal, the LVM  maps the optical stellar continuum and line emission from the ionized ISM, including dense nebulae (e.g. H~II regions, supernova remnants, and planetary nebulae) and the warm interstellar medium (WIM), across the complete disks of galaxies at very high physical spatial resolution. The LVM data provides access to key diagnostics of the physical conditions of stars and ionized gas from the small sub-pc ``energy injection scales'' in which individual ionizing sources act on the surrounding gas, up to the kpc ``galactic scales'' on which the integrated effect of these processes establishes the global thermodynamic, chemical, and kinematic structure of star forming galaxies. 

The LVM survey obtains deep, ultra-wide area, high physical spatial resolution, optical integral-field spectroscopy across the southern Milky Way (MW) plane, the Large Magellanic Cloud (LMC), and the Small Magellanic Cloud (SMC). The survey also contains two secondary programs: one covering a sparse grid of WIM-dominated high Galactic latitude pointings, and another covering a sample of nearby galaxies in the Local Group and beyond. A complete overview of the LVM, including the project's scientific motivation and experimental design, is presented in \cite{drory24}. 

A new custom-built robotic facility called the LVM Instrument \citep[LVM-I,][]{konidaris24} has been deployed at Las Campanas Observatory (LCO)\footnote{LCO is owned and operated by the Carnegie Science Observatories.} in Chile to carry out the survey. The \mbox{LVM-I} wavelength range covers all optical nebular emission lines typically used as diagnostics of the physical conditions in $\sim10^4$~K ionized gas, and the full optical stellar continuum spectrum of stars and clusters. This spectral coverage, combined with the LVM-I integral-field-unit (IFU) ultra-wide $0.5^{\circ}$ diameter field-of-view, and with the very nearby nature of our targets, allows us to map stellar populations and the ionized ISM over large areas on the sky at very high physical spatial resolution. 


A first demonstration of the science applications of the LVM data in the Orion star forming region is presented in \cite{kreckel24}. The breadth of science enabled by the LVM project is demonstrated in recent studies of the kinematical structure and physical conditions of HII regions \citep{villa-durango25, sattler26}, the detectability of Wolf-Rayet stars in low metallicity environments \citep{gonzalez-tora25}, the study of collisionless shocks in supernova remnants \citep{sarbadhicary26}, the mapping of ultra-faint metal recombination lines in HII regions \citep{Singh26}, studies of the chemical structure of planetary nebulae \citep{orozco26}, the identification of ionized gas structures towards the Galactic center \citep{kreckel26}, the characterization the stellar populations and internal kinematics of globular clusters \citep{haberle26}, and the implementation of new and powerful spectrospatial forward modeling techniques to extract weak underlying emission fields in low S/N IFU data \citep{hilder26}.

In this article we present a technical overview of the LVM project. This includes high level descriptions of the instrument, the definition and tiling of the survey footprint, the observing strategy, and the handling of the LVM data through acquisition, reduction, calibration, and analysis, in order to produce the final LVM data products. In-depth descriptions of these processes are provided in separate publications, including the Data Reduction Pipeline \citep[DRP,][]{mejia25}, the spectro-photometric calibration methodology \citep{egorova25}, the LVM approach to sky subtraction \citep{jones25}, and the Data Analysis Pipeline \citep[DAP,][]{sanchez25}. This article serves as a first stop for users aiming to use the LVM data for scientific analysis. We refer the reader to the articles mentioned above for more detailed descriptions of the methods used to acquire and process the LVM data, and to \cite{drory24} for an overview of the LVM project scientific goals and experimental design. 

A small subset of the SDSS-V LVM data has already been publicly released in the DR19 and DR20 SDSS data releases \citep{SDSSCollaboration26a, SDSSCollaboration26b}. These publicly available data include observations of a few Galactic HII regions (the Rosette Nebula, the Trifid Nebula, and the Flame and Horsehead Nebulae in Orion), a Galactic planetary nebula (the Helix Nebula) and two local volume galaxies (M33 and NGC 4945).

In section \ref{sec:instrument} we briefly describe the LVM-I, highlighting the instrument characteristics that are relevant for the analysis and interpretation of the data. This is followed by a description of the LVM survey footprint and tiling strategy in section \ref{sec:footprint}. The observing strategy is presented in section \ref{sec:observing}, followed by a description of the data handling, reduction, and calibration processes in section \ref{sec:data}. In section \ref{sec:products} we present the analysis pipeline used for spectral fitting and the production of the LVM survey data products. An assessment of the instrument stability and the quality of the data is presented in section \ref{sec:data_quality}. Finally, conclusions are given in section \ref{sec:conclusions}.

%% file: 2-instrument.tex
\section{The LVM Instrument (LVM-I)} \label{sec:instrument}

\begin{figure*}[t]
\gridline{\fig{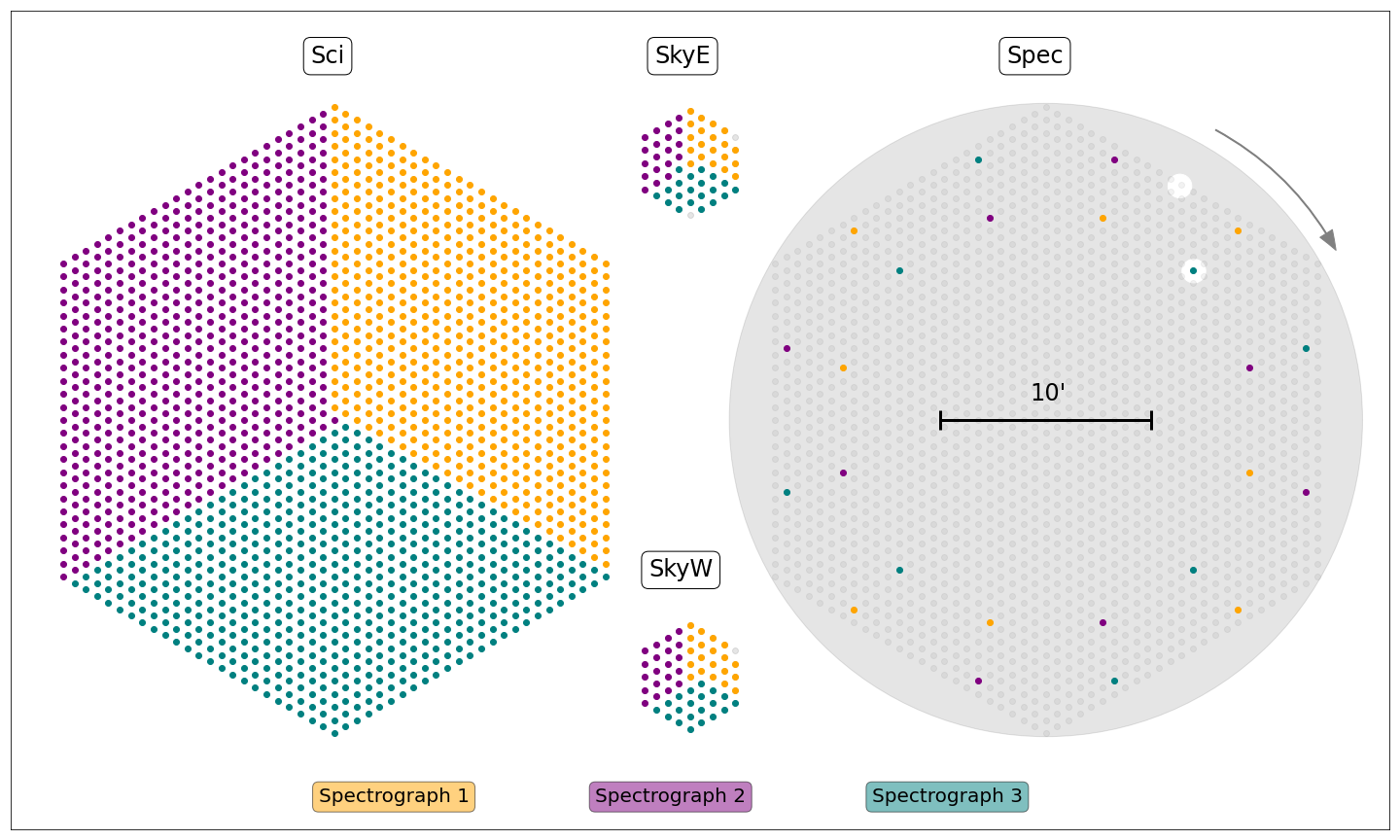}{1.0\textwidth}{}}
\vspace{-1.5\baselineskip}
\caption{
Layout of the four LVM-I IFUs that are installed at the center of the telescopes' focal surfaces. Each small circle corresponds to the center of a spaxel, color-coded by the spectrograph to which its fiber is routed, and on which its spectrum is imaged. Spaxels colored in gray are not coupled to optical fibers and no data is recorded for them. The rotating shutter used to coordinate the sequential observations of spectrophotometric standard stars during each science exposure is shown overlaid in front of the Spec telescope IFU. The labels in the figure indicate the corresponding telescopes and spectrographs.
}
\vspace{0.5\baselineskip}
\label{fig:IFUs}
\end{figure*}

The LVM-I is a dedicated standalone facility recently built and deployed at LCO in the southern part of the Atacama desert in Chile. A detailed overview of the facility and its subsystems is presented in \cite{konidaris24}. The facility is composed of five subsystems: the telescopes, the IFU fiber system, the spectrographs, the enclosure, and the instrument control software. Here we describe the first four of these sub-systems, whose properties are relevant for understanding the characteristics of the LVM data. The instrument control software is described in \cite{sanchez-gallego24}.

\subsection{Telescopes}
The LVM-I collects light using four separate 16.1~cm aperture telescopes \citep{herbst24}. One telescope (called ``Sci'' for ``science'') feeds the main science IFU, two other telescopes (``SkyE'' and ``SkyW'' for sky East and West, respectively\footnote{East and West here refer to the physical position of these units with respect to the science telescope on the telescope platform, and not to where they point on the sky.}) target blank fields\footnote{The MW WIM covers the entire celestial sphere so there is no such thing as a blank field devoid of MW nebular emission on the sky. In this context, "blank sky fields" refer to fields with the faintest levels of MW emission as discussed in Section \ref{sec:skyobs}.} and are used to obtain sky spectra, and a fourth telescope (``Spec'' for ``spectrophotometric'') targets spectrophotometric standard stars used for flux calibration and atmospheric telluric-absorption corrections. The need for separate telescopes arises from the large angular size of our targets and the low on-sky surface density of bright spectrophotometric standard stars. These two factors prevent us from obtaining sky and spectrophotometric calibration spectra within the field-of-view of an individual telescope.

The four telescopes are composed of a siderostat that delivers a non-magnified image of the sky to a static bench-mounted refractor telescope fitted with a K-mirror field de-rotator. This setup delivers a non-rotating magnified image of the sky at the static focal surface of the telescope. All four telescopes are identical in their optical design, with the exception of the Spec telescope which observes individual point sources and therefore does not require a field de-rotator. Instead, the Spec telescope is fitted with a rotating mask that can expose specific fibers to single stars in the field. This design was adopted to avoid the movement of the IFU optical fibers, and the variability in fiber throughput and line-spread-function (LSF) shape induced by those motions while tracking science fields and obtaining calibrations \citep[e.g.][]{drory15, law21}. 

The LVM-I telescopes deliver a 1.4$^{\circ}$ diameter field-of-view (FOV) fed at f/11.4, and a 0.112$\arcsec /  \mu$m plate scale \citep{lanz22}. The telescopes' FOVs are oversized with respect to the IFUs FOVs as they also need to accommodate the acquisition and guiding cameras (see Figure 13 in \cite{herbst24} and Section \ref{sec:agc} below). The image quality of the telescopes translates into an average 2.6$\arcsec$ PSF FWHM across the focal planes \citep{herbst24}. This PSF is an order of magnitude smaller than the spaxel size of the IFUs (see next section) and therefore largely undersampled.

\subsection{IFU Fiber System} \label{sec:instrument-IFU}

The IFU fiber system delivers the light from the telescopes to the LVM-I spectrographs \citep{feger22}. At the focal plane of each telescope, light is fed into an IFU composed of a pair of micro-lens arrays (MLAs) coupled to an input bundle of optical fibers. The MLAs simultaneously act as an IFU, slicing the field into hexagonal spaxels, and as a focal reducer, speeding up the f/11.4 telescope beam to feed the fibers at f/3.7. From each IFU the fiber bundle is routed to a sorting box (dubbed the ``sorting hat'') that reorganizes the individual fibers (1944 in total) into three separate output bundles. Each of the output bundles carries a combination of science, sky, and spectro-photometric calibration fibers, and terminates in a pseudo-slit configuration that feeds 648 fibers to each of the three LVM-I spectrographs. 

\begin{figure}[t]
\gridline{\fig{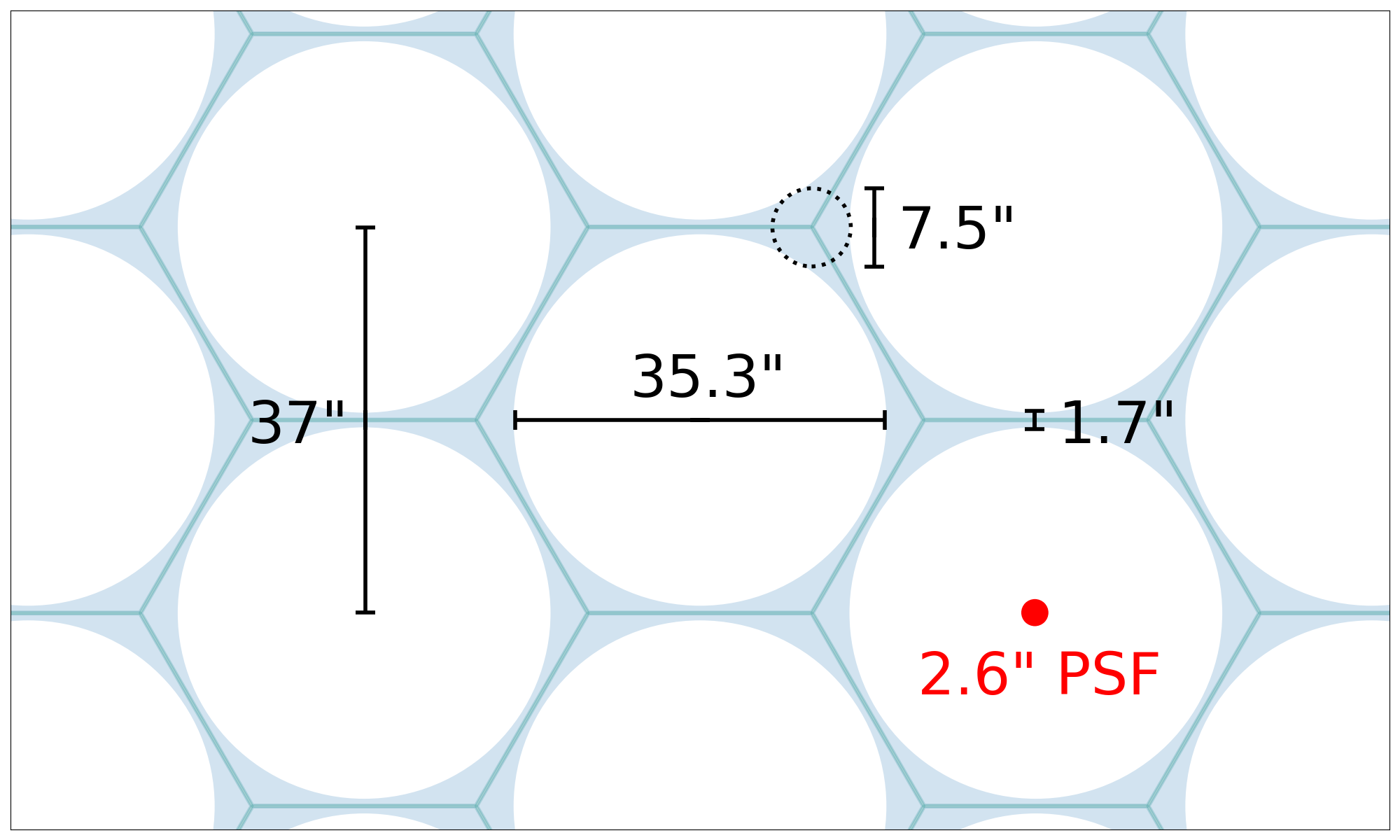}{0.47\textwidth}{}}
\vspace{-1.5\baselineskip}
\caption{Diagram showing the shape, dimension, and tiling pattern of the spaxels in the LVM-I IFUs. Each spaxel has a 35.3$\arcsec$ circular clear aperture created by the chromium mask used to block the interstitial regions between adjacent lenslets in the array. These circular spaxels are tiled following a hexagonal pattern with a 37$\arcsec$ pitch. The red circle shows a 3$\arcsec$ PSF corresponding to the average image quality delivered by the telescope optics. The chromium masks block 17\% of the area, and causes few arcsecond sized gaps in the FOV behind which point sources can be partly and even completely occulted.
}
\vspace{0.5\baselineskip}
\label{fig:spaxels}
\end{figure}

Figure \ref{fig:IFUs} presents a layout of the four IFUs in the system and the way in which fibers are distributed to different spectrographs. The science IFU has 1801 fibers in a 25-ring hexagonal tiling pattern. To prevent diffraction effects at the edges of the micro-lenses, and the associated increase in scattered light, a circular chrome mask covers the interstitial regions at their boundaries. This translates into the IFU having circular spaxels which are 35.3$\arcsec$ in diameter, arranged in a hexagonal tiling pattern with a 37$\arcsec$ pitch and a 83\% fill factor. The diagram in Figure \ref{fig:spaxels} shows the arrangement of the individual spaxels, their pitch, and the dimensions of the clear and obstructed apertures. The 2.6$\arcsec$ PSF FWHM delivered by the telescopes is also shown. In the LVM data the on-sky PSF is largely undersampled. This is a different regime than in other IFU datasets like MaNGA \citep{law15}, SAMI \citep{sharp15}, and CALIFA \citep{sanchez23} in which the PSF is either sampled or marginally sampled by the spaxels.

The on-sky footprint of the science IFU is a hexagon with 15.4$\arcmin$ (925$\arcsec$) long sides\footnote{Measured as the distance between the centers of spaxels in adjacent vertices of the hexagon.}, and a 30.2$\arcmin$ (1813$\arcsec$) outer diameter \footnote{Measured as the distance between the centers of spaxels in opposite vertices of the hexagon.}. This translates into an ultra-wide 0.165 deg$^2$ FOV that enables the efficient mapping of large areas on the sky. The sky telescope IFUs have an identical design except for their size. They are 5-ring hexagonal arrays composed of 61 lenslets. The SkyE and SkyW IFUs only have 59 and 60 spaxels populated with fibers, respectively (notice the grey spaxels in Figure \ref{fig:IFUs}). 

The spectrophotometric telescope IFU array is identical to the science one, but only 24 lenslets are populated with fibers (12 in the 19th ring and 12 in the 24th ring, as seen in Figure \ref{fig:IFUs}). This translates into 24 sparse spaxels distributed in two circles. A rotating mask in front of the IFU allows for individual spaxels to be exposed independently, and can also shutter all spaxels at the same time. This design allows the LVM-I to sequentially observe multiple bright spectrophotometric standard stars near the science field throughout the duration of each science exposure.

\subsection{Acquisition and Guiding Cameras}
\label{sec:agc}

Each telescope is fitted with a pair of CMOS acquisition and guiding cameras (AGCs) fed by two pick-off mirrors bracketing the central IFU, except for the Spec telescope, which is fitted with only one camera sampling only one side of the focal plane \citep[Figure 13 in][]{herbst24}. Each AGC covers a $26.8'\times 18.4'$ FOV, sufficiently large to detect enough stars for the purpose of acquisition, guiding, and the calculation of astrometric solutions practically anywhere on the sky. 

The relative position and rotation in the focal plane between each AGC camera and the corresponding IFU has been measured accurately in the laboratory and validated on-sky. This allows us to use the positions of stars in the AGC images to solve the astrometric solutions of the IFUs and calculate accurate on-sky coordinates for each spaxel (see Section \ref{sec:astrometry}). Stars in the AGC are also used to monitor the atmospheric transparency and make on-the-fly decisions regarding weather conditions during the observing process.

\subsection{Spectrographs}
\label{sec:spectrographs}

The three LVM-I spectrographs are fundamentally copies of the Dark Energy Spectroscopic Instrument (DESI) spectrographs \citep{edelstein18, desi22}, with some design modifications described in \cite{konidaris24} and \cite{drory24}. Each of the three units images the 648 fibers in its pseudo-slit through two nested dichroics that split the light into three wavelength channels: ``blue'', ``red'', and ``infrared'', hearafter referred as the $b$, $r$, and $z$ channels. In each channel, light is dispersed through transmission volume-phase holographic (VPH) gratings and the fiber spectra are imaged through a camera onto STA-4850 4k$\times$4k CCDs with 15$\mu m$ pixels. The $z$ channel detectors are a thick deep-depletion variant of the same device, which improves their quantum efficiency and prevents fringing effects. 

\begin{figure}[th]
\gridline{\fig{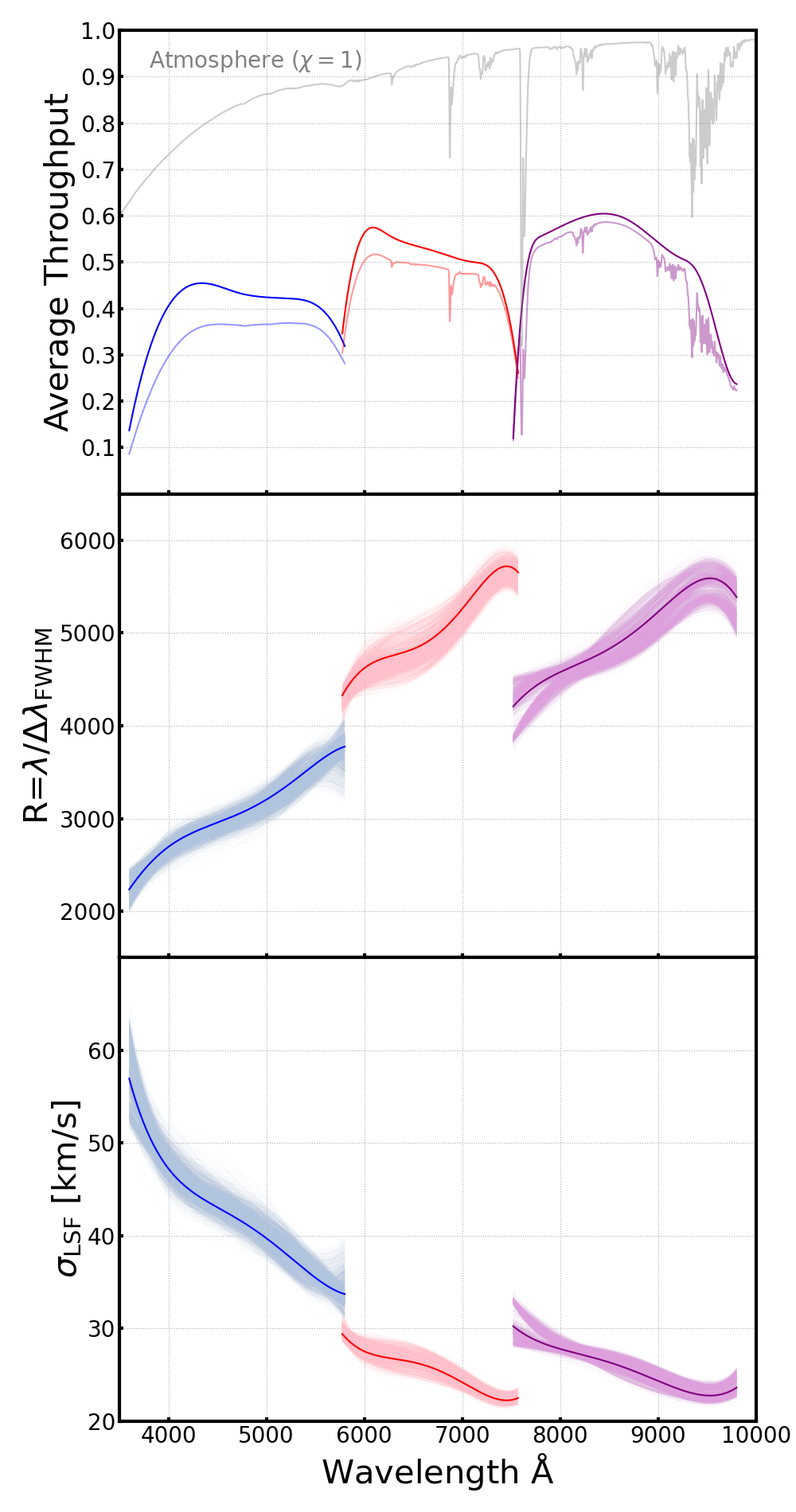}{0.47\textwidth}{}}
\vspace{-2\baselineskip}
\caption{
\textit{\textbf{Top Panel:}} LVM-I average system throughput measured on-sky (lighter solid lines) for the $b$, $r$, and $z$ channels (in blue, red, and purple respectively). The darker lines show the total \mbox{LVM-I} throughput including atmospheric extinction at one airmass (grey line).  
\textit{\textbf{Middle Panel:}} Spectral resolution of the LVM-I fibers as a function of wavelength. Dark solid lines show the average resolution in each channel across all 1944 fibers.
\textit{\textbf{Bottom Panel:}} Instrumental LSF velocity resolution in km~s$^{-1}$. Small changes to LSF of each fiber happen whenever the spectrograph cameras are re-focused. The LVM data products report the applicable instrumental LSF FWHM as a function of wavelength for each fiber in each exposure. 
}
\vspace{0.5\baselineskip}
\label{fig:channels}
\end{figure}

The gratings, camera glasses, coatings, and detectors are optimized to maximize the efficiency of each channel. Figure \ref{fig:channels} presents the throughput and spectral resolution of the LVM-I spectrographs. The wavelength ranges of the three channel are $b$: 3600-5800~\AA,  $r$: 5775-7570~\AA, and $z$: 7520-9800~\AA. The spectrographs deliver average spectral resolutions ($R=\lambda/\Delta\lambda_{FWHM}$) of $R=2900$, 4690, and 4870 in the blue, red, and infrared channels respectively. The average instrumental line-spread-function width ($\sigma_{LSF}$) of each channel covers a range of 34-57~km~s$^{-1}$ in the $b$-channel, 22-29~km~s$^{-1}$ in the $r$-channel, and 23-30~km~s$^{-1}$ in the $z$-channel.

Along the output pseudo-slit of each spectrograph, fibers are arranged in 18 blocks of 36 fibers each \citep[i.e 54 blocks hosting a total of 1944 fibers across the  three spectrographs,][]{feger22}. The 119 fibers coming from the SkyE and SkyW telescopes (Section \ref{sec:instrument-IFU}) are routed so the top and bottom fiber of each block is a sky fiber. The remaining 11 sky fibers populate the second from the top and second from the bottom position of a subset of the blocks. The 24 fibers coming from the Spec telescope are routed so eight blocks in each spectrographs recieve one calibration fiber. Since we typically observe 12 spectro-photometric standard stars during a science exposure (see Section \ref{sec:observing}), each spectrograph typically images the spectra of four calibration stars across the three channels. The bulk of the fibers in each block (33 to 34 per block) are science fibers coming from the Sci telescope. Figure \ref{fig:IFUs} shows how different areas within each IFU are mapped to the three LVM-I spectrographs.

Along the spatial direction the fiber profile has an average FWHM of 2.5~pixels. Adjacent fibers inside each block ar spaced by $\sim$5.5~pixels, which translates into a low ($\simeq$0.16\%) level of flux cross-talk between adjacent fibers. This cross-talk \footnote{Cross-talk is defined as the fractional area of a fiber PSF that is overlapped by that of an adjacent
fiber.} is modeled and accounted for during the extraction of the fiber spectra (see Section \ref{sec:drp}).

\subsection{Enclosure and Calibration System}
\label{inst:cal}

The LVM-I is hosted inside a roll-off roof enclosure that covers the telescope platform \citep{blanc24}. The enclosure also incorporates an environmentally-controlled and clean room for the spectrographs, plus space for servers, network, HVAC, detector cooling, and power utilities. The roll-off design allows the four telescopes to simultaneously access the whole sky visible above an airmass of $\chi=2$. 

An external calibration system with two types of continuum flat-field sources: a Laser Driven Light Source (LDLS) and a quartz-tungsten-halogen (QTH) lamp; and four types of arc lamps (Ar, Ne, HgNe, and Xe); is installed on the telescope platform. The calibration lamps illuminate a flat screen attached to the inside of the roll-off roof. Taking calibrations requires the dome to be closed and all telescopes to be pointed at the calibration screen.

The thermal stability of the spectrographs has a direct impact on the calibration strategy of the survey. Changes in temperature of $\sim1^{\circ}$C can induce $>0.5$ pixel level shifts in the positions of the fiber spectra on the detector, with an associated impact on survey efficiency because of the potential need to obtain calibrations at a high cadence. Keeping the spectrographs in a thermally stabilized environment at $20\pm0.2^{\circ}$C minimizes thermal effects, and in combination with our static optical fiber system design, minimizes overheads associated with obtaining calibrations \citep{blanc24}.

%% file: 3-footprint.tex
\section{Survey Footprint and Tiling Strategy} \label{sec:footprint}

\begin{figure*}[t]
\gridline{\fig{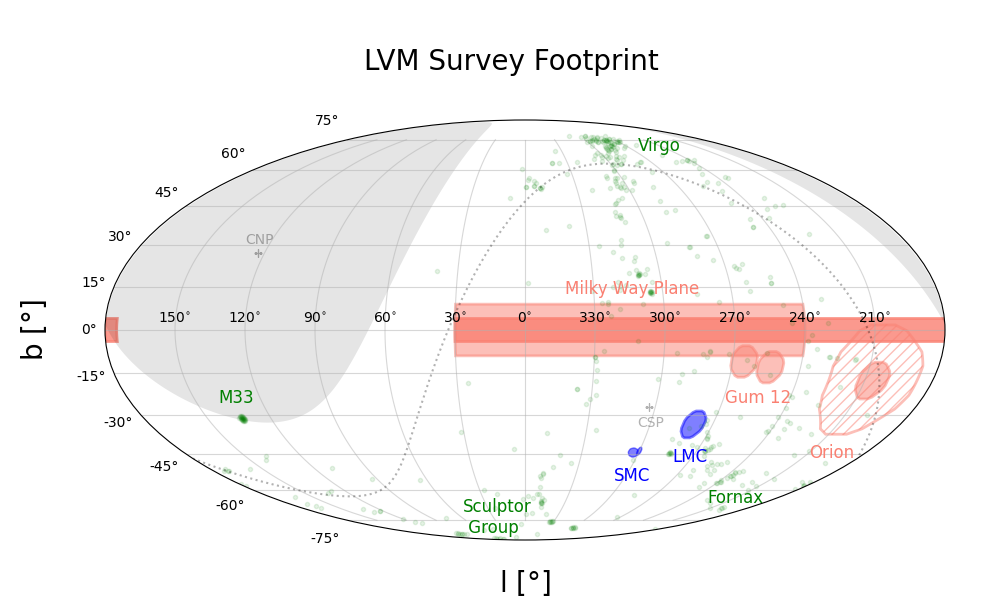}{1.0\textwidth}{}}
\vspace{-1.5\baselineskip}
\caption{LVM survey footprint on an all-sky Mollweide projection in Galactic coordinates. The gray crosses and dotted line mark the celestial poles and the equator, respectively. The solid gray area marks northern declinations that are not accessible at airmass $\chi<2$ from LCO. MW targets are shown in red, with the high priority 8$^{\circ}$ tall band along the Galactic mid-plane in a darker tone than the 8$^{\circ}$ tall extension band. Also visible are the low Galactic latitude areas that target the Gum nebula and the Orion star forming region. The outer hatched areas in Orion is sparsely sampled. The Magellanic Cloud coverage is shown in blue and the Local Volume galaxy sample coverage is shown in green. For clarity we do not show the MW high Galactic latitude sparse grid, but these cover the full white area of the diagram with a 1/90th fill factor.
}
\vspace{0.5\baselineskip}
\label{fig:footprint}
\end{figure*}

The LVM core targets are the Milky Way and the Magellanic Clouds. Within the MW, the main focus of the survey is on star forming regions in the Galactic plane. We also target the Orion star forming region and the Gum nebula (a.k.a. Gum-12), whose proximity to the Sun ($D\simeq400$~pc) makes them appear at higher Galactic latitudes than the main mid-plane survey footprint. When the core targets are not available, the LVM observes a sample of nearby galaxies in the local volume and a sparse grid of high Galactic latitude fields to map the MW extra-planar diffuse ionized gas (DIG) emission. The latter two target groups serve as a filler program and are not ensured to be completed during the survey. The areas covered by the LVM are listed in Table \ref{tab:targets} and shown in Figure \ref{fig:footprint}. A detailed presentation of the scientific motivation and definition of the survey footprint is presented in \cite{drory24}. Here, we provide a brief summary of the targeted areas and describe the strategy used to segment the footprint into individual tiles that can be covered by the LVM-I science IFU.

\input{t1_tab_targets}

\subsection{Milky Way}

{\it - Galactic Plane:} we target a high priority $8^{\circ}$ tall and $215^{\circ}$ wide band that covers the $-4^{\circ}<b<4^{\circ}$ range in galactic latitude, and the bulk of the galactic longitude range accessible from LCO. That is, starting from $l=30^{\circ}$, passing through the Galactic center, and extending out to $l=175^{\circ}$ (i.e. past the MW anti-center, see Figure \ref{fig:footprint}). A lower priority $-9^{\circ}<b<9^{\circ}$ extension of this band, with a more limited range in Galactic longitude ($l<30^{\circ}$ and $l>240^{\circ}$) is also targeted. In the MW, the LVM-I spaxel angular size corresponds to a physical spatial resolution of $<1$~pc out to a distance $D<6$~kpc.\\

{\it - Orion Region:} the survey footprint includes a circular area centered on the Orion star forming region. Contiguous coverage is provided for a $6.5^{\circ}$ radius area centered at $b=-17.74^{\circ},l=206.42^{\circ}$.  This area encompasses the Orion nebula (M42/M43), the Horsehead nebula (Barnard 33) and its background HII region (IC432), the Flame nebula (Sh2-277), and the bulk of Barnard's Loop (Sh2-276). Beyond this radius a sparse grid of tiles with a 1/5$^{\rm{th}}$ filling factor samples the Orion-Eridanus super-bubble out to a maximum radius of $19.5^{\circ}$ ($\sim270$~pc in diameter). In Orion \citep[$D\simeq400$~pc,][]{kounkel18} the LVM-I IFU reaches a spatial resolution of $\simeq0.07$~pc per spaxel.\\

{\it - Gum Nebula (Gum 12):} the survey footprint includes two circular areas with $5.5^{\circ}$ radius centered at $b=-13.00^{\circ},l=253.20^{\circ}$ and $b=-11.00^{\circ},l=265.00^{\circ}$ that extend the Galactic plane coverage to include the SW limb of Gum 12 (including its cometary globules) and the IRAS Vela shell, respectively. For Gum 12 \citep[$D\simeq500$~pc][]{woermann01}, the LVM-I resolution is $\simeq0.09$~pc per spaxel.\\

{\it - High Galactic Latitude Sparse Grid:} the survey also includes an all-sky regular grid of pointings which are sparsely sampled with a 1/90$^{\rm{th}}$ filling factor. This sparse grid provides the ability to map nebular emission from extra-planar diffuse ionized gas at high Galactic latitudes. It also serves as a default database of low priority filler targets for the overall survey.

\subsection{Magellanic System}

{\it - Large Magellanic Cloud:} the LMC is covered by a $5^{\circ}$ radius circular area ($\sim9$ kpc in diameter) centered at equatorial coordinates ($\alpha,\delta)=(79.51^{\circ},-68.53^{\circ}$). This area covers practically the entire galaxy out to its optical semi-major(minor) axis $R_{25,LMC}=5.4^{\circ}$(4.6$^{\circ}$), \citep{devaucouleurs91}. At the distance of the LMC  \citep[D$\simeq49.6$~kpc][]{pietrzynski19}, the LVM-I spaxels resolve scales of $\simeq8$~pc.\\

{\it - Small Magellanic Cloud:} coverage of the SMC is achieved by a primary elliptical area centered at ($\alpha, \delta)=(13.16^{\circ},-72.80^{\circ}$), with semi-major and minor axes $(a,b)=(2.8^{\circ}, 1.6^{\circ})$ ($\sim 7$~kpc across), and a position angle $PA=45^{\circ}$. This area covers the main body of the galaxy beyond its optical semi-major(minor) axis $R_{25,LMC}=2.6^{\circ}$(1.6$^{\circ}$), 
\citep{devaucouleurs91}. We extend this footprint using a second elliptical area centered at ($\alpha, \delta)=(20.00^{\circ},-73.20^{\circ}$), with semi-major and minor axes $(a,b)=(1.5^{\circ}, 1.0^{\circ})$, and a position angle $PA=155^{\circ}$. The later covers the SMC Wing structure, a plume of gas and associated star formation that extends up to 2~kpc SW of the galaxy \citep{fulmer20}. At the distance of the center of the SMC  \citep[$D\simeq62.4$~kpc][]{graczyc20}, the LVM-I achieves a physical spatial resolution of $\simeq11$~pc per spaxel.

\subsection{Local Volume Galaxies}

{\it - Local Group Sample:} we target 29 Local Group star forming galaxies with distances spanning the 0.5--5.6~Mpc range. These targets are listed in Table 2 of \cite{drory24}, and they are accessible from LCO ($\delta\le15^{\circ}$, except for M33 at $\delta=+30.7^{\circ}$ for which we allow an exception to our airmass limit). They are covered by single LVM-I pointings, with the exception of M33 (24 pointings) and NGC 55, NGC 247, NGC 253, NGC 300, NGC 1313, NGC 4945, and NGC 5128/CenA (each covered by 7 pointings). Over the distance range of the Local Group targets the LVM-I spatial resolution ranges from 90~pc to 1~kpc per spaxel. This is an intermediate spatial resolution regime between those achieved by the PHANGS-MUSE \citep{emsellem22} and the CALIFA \citep{sanchez12, sanchez23} surveys. \\

{\it - 20~Mpc Sample:} we extend the Local Volume galaxy sample with a list of $\simeq400$ massive ($M_*\ge10^9\rm{M}_{\odot}$) star forming galaxies out to a distance $D<20$~Mpc \citep[see Section 5.3.2 in][]{drory24}. A single LVM-I pointing covers the entire optical disk (beyond $R_{25}$) for all these targets. For these galaxies the IFU spaxels sample spatial scales in the 1.0-3.5~kpc range, depending on target distance. This is similar to the spatial resolution range of the MaNGA survey \citep{bundy15}.


\subsection{Tiling and Dithering Strategy}
\label{sec:tiles}

To observe these targets, the areas described above need to be segmented into hexagonal tiles the size of the LVM-I science IFU. Individual tiles then become the fundamental units that are observed with the LVM-I, and they are also the basic dataset units used to store, handle, and distribute the reduced LVM data (in a similar way to how plug-plates were treated in previous SDSS data-models).

\input{t2_tab_dithers}

A tile is defined by its center coordinates and its position angle (PA)\footnote{In the LVM, PA=0 corresponds to an IFU rotation with two opposing flat sides of the hexagon facing North and South, and two opposing vertices of the hexagon pointing East and West.}. To segment the survey footprint into individual tiles, we first create a set of bounding polygons that encompass all targets whose area shares some level of overlap (e.g. all MW targets as seen in Figure \ref{fig:footprint}). These bounding polygons are tiled independently, allowing us to generate an optimized contiguous tiling pattern for different disjoint areas covered by the survey. For each of these bounding polygons, we draw a great circle that crosses the area-weighted bary-center of the polygon along its longest direction, and define a line of slightly overlapping tiles following a hexagonal pattern along this great circle. The pitch of the tiling pattern is such that neighboring tiles have approximately 2 spaxels of overlap along their adjacent sides. Once this line of tiles is defined, we grow the tiling pattern perpendicularly above and below the previously mentioned great circle until the complete bounding polygon area is covered.

In the case of sparsely sampled fields like the outer regions of Orion and the high Galactic latitude MW grid, we define the full tiling pattern in the same way as above, and then flag a sparse sub-sample of tiles for observation. This will allow filling in the coverage of these areas in the future if desired.

All tiles corresponding to MW targets are observed with no dithering, pointing the science IFU at the defined tile coordinates. This introduces gaps in the spatial coverage of these areas due to the 83\% fill factor of the IFU (see Figure \ref{fig:spaxels} and Section \ref{sec:instrument-IFU}). These gaps in the coverage can be eliminated by obtaining dithered observations, but in the case of the MW survey, this is unfeasible given the survey duration and the area we aim to cover during the duration of the SDSS-V project. 

\begin{figure}[t]
\gridline{\fig{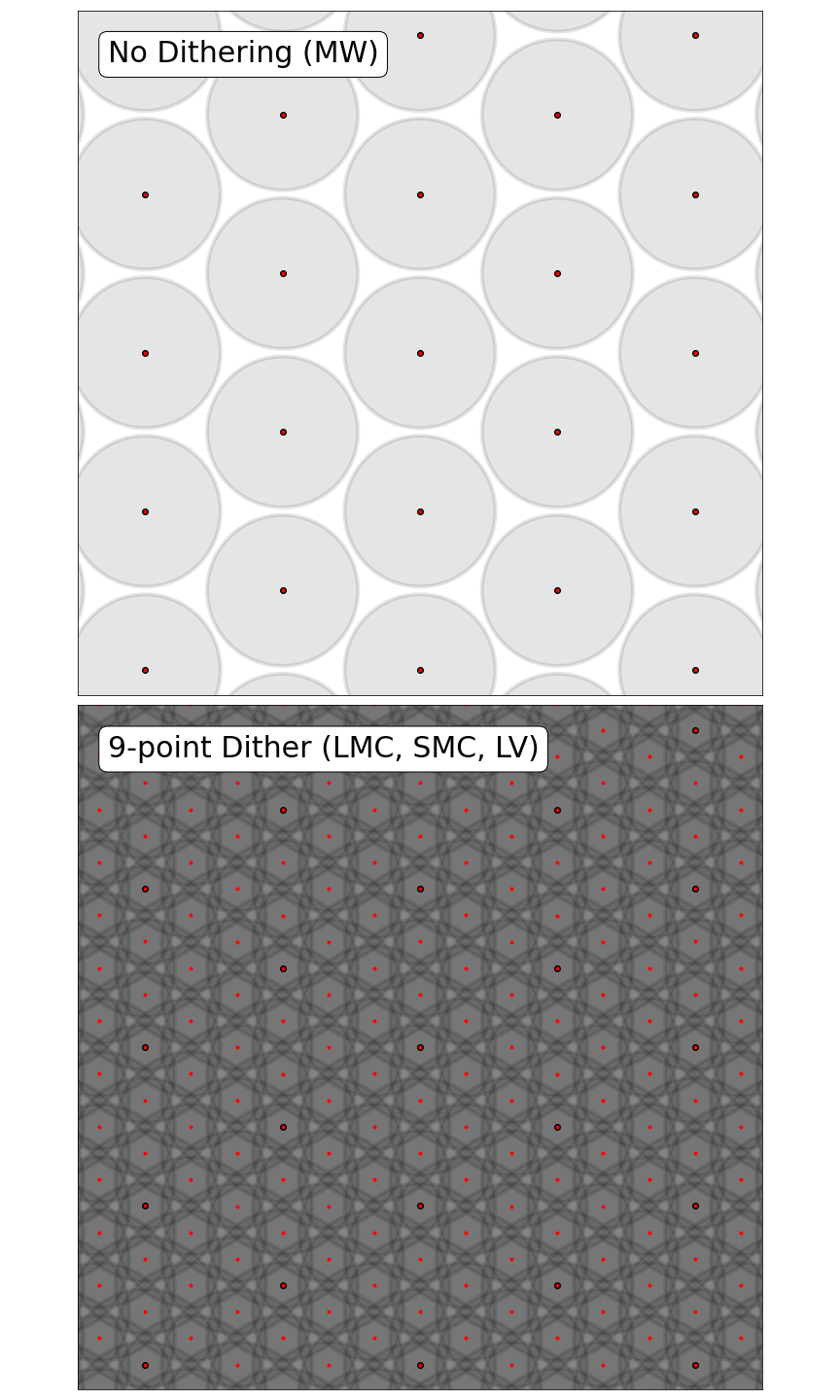}{0.5\textwidth}{}}
\vspace{-1.5\baselineskip}
\caption{
Comparison of the aerial on-sky coverage of the LVM-I science IFU spaxels when doing a single non-dithered observation as adopted in the MW survey (top panel), versus the nine-point dithered observation strategy adopted for the Magellanic Clouds and the Local Volume galaxy sample. The gray apertures correspond to the 35.3$\arcsec$ aperture of each spaxel. The dark red dots in both panels correspond to the centers of the spaxels when pointing at the fiducial tile coordinates. Light red dots in the bottom panel show the centers of the same spaxels in the nine dithered positions.
}
\vspace{0.5\baselineskip}
\label{fig:dither}
\end{figure}

For tiles in the Magellanic Clouds and the Local Volume galaxy sample, we do adopt a dithering strategy that ensures uniform coverage of these targets with no gaps. In these cases, each tile is observed nine times applying small dither offsets in between exposures. Table \ref{tab:dithers} lists the magnitude of the nine dithered offsets in arcseconds with respect to the center position of the tile, and Figure \ref{fig:dither} shows a comparison of the aereal coverage obtained in the MW using single exposures and in external galaxies using nine dithered exposures. When using this dither pattern, 100\% of the observed area is sampled by spaxels in at least 6 exposures. The percentage of the observed area covered by 6, 7, 8, and 9 exposures is 8.5\%, 46.7\%, 41.3\%, and 3.4\% respectively (see bottom panel of \ref{fig:dither}).

Table \ref{tab:targets} lists the different targets included in the survey footprint, their effective surveyed area, the number of tiles and dithers use to observe them, the inter-tile and intra-tile filling factors corresponding to the adopted observing strategy for each target, and their corresponding total exposure time per tile. Overall, the survey covers 4,598 deg$^2$ across the full footprint, corresponding to 28,927 LVM-I science IFU tiles, and $\sim68$ million independent spectra\footnote{Approximately 17\% of this area corresponds to the filler targets (local volume galaxies and MW high galactic latitude sparse grid) which are not ensured to be completely observed during the survey.}.

%% file: t1_tab_targets.tex
\begin{deluxetable*}{cccccccc}
\tabletypesize{\footnotesize}
\tablecolumns{7}
\tablewidth{0pt}
\tablecaption{ LVM Survey Targets and Tile Coverage \label{tab:targets}}
\tablehead{
\colhead{Target} & \colhead{Area\tablenotemark{\scriptsize a}} & \colhead{$N_{Tiles}$} & \colhead{Inter-Tile}  \vspace{-0.3cm} & \colhead{$N_{Dithers}$} & \colhead{Intra-Tile} & \colhead{Exposure Time\tablenotemark{\scriptsize d}} &
\colhead{Priority} \\ 
& [deg$^2$] &  & \colhead{Filling Factor\tablenotemark{\scriptsize b}} &  & \colhead{Filling Factor\tablenotemark{\scriptsize c}}  & [s]} 
\startdata
\hline
\hline
Milky Way & 4,421 & 27,817 & & & & & \\
\hline
Galactic Plane (main) & 1,789 & 11,249 & 1 & 1 & 0.826 & 900 & 4 \\ 
Galactic Plane (extension) & 1,562 & 9,826 & 1 & 1 & 0.826 & 900 & 6 \\ 
Orion (center) & 136 & 854 & 1 & 1 & 0.826 & 900 & 2 \\ 
Orion (sparse) & 36 & 229 & 1/5th & 1 & 0.826 & 900 & 3 \\ 
Gum 12 & 194 & 1,217 & & & & & 5 \\
MW High Galactic Latitude & 704 & 4,442 & 1/90th & 1 & 0.826 & 900 & 9 \\ 
\hline
\hline
Magellanic System & 98 & 616 & & & & & \\
\hline
LMC & 81 & 508 & 1 & 9 & 1.0 & 8,100 & 1 \\ 
SMC & 17 & 108 & 1 & 9 & 1.0 & 8,100 & 1 \\ 
\hline
\hline
Local Volume Galaxies & 79 & 494 & & & & & \\
\hline
Local Group Sample & 15 & 94 & 1 & 9 & 1.0 & 8,100 & 7 \\ 
20~Mpc Sample & 64 & 400 & 1 & 9 & 1.0 & 8,100 & 8 \\ 
\hline
\hline
TOTAL LVM Survey & 4,598 & 28,927 &  &  &  & &  \\ 
\enddata
\tablenotetext{a}{Area covered by tiles (i.e. ignoring intra-tile filling factor).}
\tablenotetext{b}{Fraction of target total area covered by tiles.}
\tablenotetext{c}{Fraction of tile area covered by unmasked spaxels area (see Section \ref{sec:instrument-IFU}).}
\end{deluxetable*}

%% file: t2_tab_dithers.tex
\begin{deluxetable}{ccc}
\tabletypesize{\footnotesize}
\tablecolumns{3}
\tablewidth{0pt}
\tablecaption{ LVM 9-point Dither Pattern\tablenotemark{a} \label{tab:dithers}}
\tablehead{
\colhead{Dither} & \colhead{\hspace{1cm}$\Delta$x\tablenotemark{b}}\hspace{1cm} & \colhead{\hspace{0.5cm}$\Delta$y\tablenotemark{b}}\hspace{0.5cm} \\ 
N & \colhead{\hspace{1cm}['']}\hspace{1cm} & \colhead{['']}\hspace{0.5cm}}
\startdata
1 & 0.00 & 0.00  \\ 
2 & -10.68 & 18.50  \\ 
3 & 10.68 & 18.50  \\ 
4 & -0.00 & -12.33 \\ 
5 & 10.68 & -6.17  \\ 
6 & -10.68 & -6.17  \\ 
7 & 10.68 & 6.17  \\ 
8 & -10.68 & 6.17  \\ 
9 & 0.00 & 12.33  \\ 
\enddata
\tablenotetext{a}{Dithering is only used for the Magellanic Clouds and the LVG sample.}
\tablenotetext{b}{Offsets are performed in the coordinate system of the Sci telescope focal plane. For a position angle $PA=0^{\circ}$ coordinates offsets in x and y correspond to negative ofsets in right ascension and positive offsets in declination.}
\vspace{-2.0\baselineskip}

\end{deluxetable}

%% file: 4-observing.tex
\section{Observing Strategy} \label{sec:observing}

The LVM-I is designed to be a robotic facility capable of carrying out the survey observations autonomously. Fully robotic operations need to ensure the safety of the personnel, the protection of the hardware, the quality of the data, and the efficiency of the survey. This requires extensive knowledge of the different failure modes of the hardware, the software, and the data acquisition strategy vis-a-vis weather conditions. Accumulating this knowledge is required before moving into robotic operations, so during the initial stages of the project observations were supervised by human remote observers. After a successful commissioning period, survey observations officially started in this remote mode on November 6th 2023. Regular human supervision was suspended on August 1st 2025, and since that date the LVM-I operates in a robotic autonomous mode. The execution of the baseline survey is envisioned to last four years through late 2027. 

\subsection{Science Observations}
\label{sec:obs-science}

The basic LVM observing unit corresponds to an individual exposure of a tile in the survey footprint (Section \ref{sec:tiles}). After a tile is chosen by the scheduling software (see Section \ref{sec:schedulling}), science exposures are executed by pointing the siderostat and rotating the K-mirror image de-rotator of the Sci telescope to the coordinates and position angle of the corresponding tile. At the same time, the SkyE and SkyW telescopes are pointed at optimally selected sky fields (see Section \ref{sec:skyobs}) that are chosen by the scheduling software, based on the tile's ephemeris and the time of the observation. The scheduling software also assigns 12 nearby spectro-photometric standard stars for the Spec telescope to target throughout the duration of the science exposure (see Section \ref{sec:standards}).

For tiles belonging to MW targets, we obtain a single 900~s exposure with no dithering. While the LVM is mainly focused on the measurement of extended nebular emission on scales larger than the IFU spaxels, it should be noted that the masked gaps between the spaxels are sufficiently large to occult compact and point sources (see Figure \ref{fig:spaxels}). This, together with the potential impact of differential atmospheric refraction (DAR) for stars near the edge of a spaxel, must be taken into account when analyzing and interpreting the LVM data (especially the stellar continuum in the spectra). The effects of DAR in particular are expected to be much smaller than is typical seen with other IFUs on larger telescopes (e.g.\ MaNGA, MUSE), as the LVM-I telescopes' $\sim3 \arcsec$ image quality significantly smears this chromatic effect. For the LMC, the SMC, and the Local Volume galaxies, we obtain nine 900~s dithered observations for a total exposure time of 2.25~hrs per tile (see Table \ref{tab:dithers} and Figure \ref{fig:dither}). As mentioned above, the dithers completely cover the gaps between spaxels and yield a 100\% intra-tile fill factor.

After a tile is selected for observation, the LVM-I executes an automated pointing, acquisition, and guiding process. During acquisition, an astrometric solution for the AG cameras' images is calculated on-the-fly using the {\tt astrometry.net} software package \citep{lang10}, and the calculated offsets are applied until the Sci telescope is within 2$\arcsec$ of its target position and the other telescopes within 3$\arcsec$. At this point, the guiding loop starts (except for the Spec telescope which does not require guiding due to the short exposure times), the telescopes quickly converge to within 1$\arcsec$ of their target positions, and the spectrograph shutters open. 

The Spec telescope starts by acquiring the first of the 12 selected spectrophotometric standard stars. After the desired exposure time for the star is obtained ($\sim$1/12th of the science exposure minus acquisition overheads), the rotating shutter moves to block all the Spec focal plane fibers. The telescope then acquires the next star on a different fiber, and the rotating shutter moves to commence the exposure of the second star. This process is repeated throughout the duration of the science observation to expose all 12 spectrophotometric standards on the detector prior to readout. Only fibers in the inner (primary) ring of the Spec focal plane are used to observe standards (see Figure \ref{fig:IFUs}). The outer (secondary) ring of fibers is considered a backup, and will only get used if fibers in the inner ring break or degrade during the survey. Typically all 12 planned standards are successfully observed, but acquisition problems can translate into $\sim$10\% of the exposures having 10--11 calibration spectra. In the rare instances for which we successfully acquire and expose fewer than 9 standard star spectra the tiles are flagged to be re-observed.

Once the science exposure is completed, the \mbox{LVM-I} acquires the next tile while the spectrograph CCDs are reading out, achieving typical overheads of $\sim80$~s between exposures ($<10\%$ of the exposure time). After readout, nine separate {\tt FITS} image files are written (one per spectrograph camera). Each image contains the 2D spectra of 648 fibers in each of three channels (b, r, and z) of each of the three spectrographs. We also create and store a coadded image for each of the seven AG cameras in the system. This is done by combining all individual guider exposures obtained during the science exposure using a median statistic with 3-$\sigma$ clipping. These coadded images contain PSF, astrometric, and photometric information that is used during the quality assessment, reduction, and analysis of the data.

\subsection{Calibrations}
\label{sec:obs-calib}

Calibration data is required for the purpose of detector detrending, flat-fielding, calculating wavelength solutions, tracing and extracting the fiber spectra, calibrating the relative sensitivity of different telescopes, fibers, and spectrographs, doing flux calibration, and subtracting the sky emission from the science spectra. Some of the LVM calibration data is obtained using the calibration system described in Section \ref{inst:cal}, and some of the data is obtained on sky, both during twilight and in parallel with the science observations. 

The LVM-I was specifically designed to achieve superb fiber tracing, wavelength solution, and LSF stability. The fixed focal plane and static fiber bundle design, together with the spectrograph enclosure temperature control, translate into into an excellent stability in these quantities from night to night (see Section \ref{sec:stability}). This allows for calibration datasets taken days, and even weeks, apart from the science observations to be used during the reduction process. 

Significant shifts in the tracing, wavelength solution, and LSF of the fibers on the detector do occur when the liquid nitrogen (LN2) cooled spectrograph cameras are thermally cycled from room to cryogenic temperatures, whenever the fiber pseudo-slit assembly is removed and reinstalled, and when the spectrographs are re-focused. Some of these activities happen as part of scheduled instrument maintenance activities every few months, and define date ranges across which calibrations cannot be assumed to be stable and applicable. All LVM calibration data is therefore tagged in a way that reflects these stability periods. Moderate earthquakes, which are common at the site, also define new calibration stability periods.


\subsubsection{Bias Frames and Pixel Flats}

CCD detrending requires the removal of thermally induced dark current, the readout electronic bias signal, and pixel-to-pixel quantum efficiency variations. For the LVM-I detectors, we have measured the dark current to be negligible over the adopted exposure times (0.001~e$^{-}$s$^{-1}$). We therefore do not take darks as part of our regular calibrations. Sets of seven bias frames are obtained on a daily basis. 

Obtaining pixel flats with the LVM-I is non-trivial, as the detectors cannot be uniformly illuminated by the pseudo-slit fibers. To overcome this problem, we physically remove the pseudo-slit assembly of each spectrograph and replace it with a pseudo-slit that is specially designed for this purpose. This ``flat-field slit" uses a combination of fibers, a cylindrical lens, and diffusers, to smooth out the high-frequency spatial structure of the pseudo-slit. In this setup we can illuminate the \mbox{LVM-I} detectors in a sufficiently uniform fashion across the spatial direction to allow the production of pixel-to-pixel flat-field frames. 

Pixel flats are taken with the two continuum lamps in the calibration system (LDLS and QTH, Section \ref{inst:cal}).
The QTH lamp provides good illumination with a smooth spectral shape in the $r$ and $z$ channels, but it is not bright enough in the blue. The LDLS lamp provides sufficient signal for flat-fielding the $b$ channel, and is free of sharp spectral features in this channel (unlike at redder wavelengths). This procedure is executed on a yearly basis, and a minimum set of 100 individual pixel flat-field frames is obtained each time.

\subsubsection{Arc Frames}

Arc lamp frames are obtained using the calibration system with the HgNe, Ne, Ar, and Xe lamps turned on simultaneously. We observe two sets of 12 short (10~s) and 12 long (50~s) exposures. Two exposure times are used because of the large dynamic range in line brightness across the arc spectra. 
In each set of 12 arc exposures we cycle the Spec telescope rotating shutter to expose one of the fibers in the inner primary ring. We obtain full sets of arcs following this procedure on a monthly basis. On a nightly basis, we obtain a single short and a single long arc exposure (expsing a signle Spec fiber at random) for the purpose of monitoring the wavelength solution stability. 

\subsubsection{Dome and Twilight Fiber Flats}

Dome flats are used for tracing the fibers on the detectors and to measure their profile shapes in the cross-dispersion direction. Two sets of 12 dome flats (cycling through the Spec telescope fibers) are obtained on a monthly basis. One set uses the QTH lamp with 20~s exposures, and the other the LDLS lamp with 150~s exposures. As discussed above, the QTH lamp is used to trace the fibers in the $r$ and $z$ channels and the LDLS lamp to trace them in the $b$ channel. On a nightly basis we obtain single QTH and LDLS flat frames to monitor the stability of the instrument. 

The relative illumination of the fibers in the dome flats differs from the on-sky illumination. This prevents us from using the dome flats to correct for differences in the relative throughput of different fibers fed from different telescopes, and being imaged through different spectrographs. Twilight flats, on the other hand, do not suffer from biases in illumination, and can be used for this purpose. On a daily basis, we obtain sets of 12 twilight flats (again cycling through the Spec fibers) with all four telescopes pointed at the same position on the sky and with telescope tracking turned off. A script automatically adjusts the exposure times to ensure an adequate signal level is maintained in the frames as twilight progresses.

\subsubsection{Spectro-photometric Standard Stars}
\label{sec:standards}

As described in Section \ref{sec:obs-science}, each science exposure is complemented by the simultaneous observation of a set of 12 spectro-photometric standard stars. These stars are selected from GAIA DR3 \citep{gaia16, gaia23}, using a criterion aimed at building a clean sample of F-type main sequence stars bright enough to reach a continuum V-band $S/N>25$ per \AA\ in $\sim75$~s of exposure in bright time. The LVM calibration spectra are therefore in a similar regime to that of the \mbox{SDSS-IV} MaNGA survey in terms of number of calibration sources and spectra S/N. This allows us to adopt a similar approach for flux calibration as the one used and validated in MaNGA \citep{yan16}.

In particular, we select stars with Gaia G-band magnitudes in the $5<G<9$ range, with effective temperatures measured from Gaia BP/RP spectra in the $6000~{\rm K}<T_{\rm eff}<7500~{\rm K}$ range \citep{gaia_apsis2023}. We further require the stars to have parallax errors of $<$20\%, no flags for stellar variability, and have published BP/RP spectra available in Gaia DR3. Accounting for our 35.3$\arcsec$ diameter spaxels, this magnitude range translates into calibration sources that are 4-150 (100-4000) times brighter than the sky in bright (dark) time. As the Galactic plane is crowded with faint stars, we minimize the impact of stellar continuum light from other nearby stars by calculating the flux within the fiber arising from all Gaia stars brighter than 21~mag in G-band, and reject stars where this integrated light is more than 1\% the brightness of the calibration star in the red RP-band. 

The final LVM calibration star catalog contains 16,482 objects. As discussed in \cite{drory24} (see their Figure 7), the on-sky surface density of selected calibration stars ($\sim0.4$~deg$^2$) is sufficiently high that we can observe a full set of 12 standards within a 3 (5) degree radius from the center of most (all) survey tiles. In rare cases in which the acquisition of some standard stars fail, we require a minimum of 9 successfully observed calibrators to consider the science observation completed. This ensures at least one standard star is observed in each spectrograph. 

\subsubsection{Sky Spectra}
\label{sec:skyobs}

The two sky telescopes are pointed to blank\footnote{There are no perfectly blank fields in the sky (as in devoid from any astrophysical emission). For our purposes, these are fileds in which emission from astrophysical origin is significantly fainter than both the atmospheric emission that is being removed and the astrophysical signal in the science field that is being recovered after sky subtraction.} sky fields whose spectra are used for the purpose of sky subtraction. \cite{drory24} describe the way in which these fields are pre-selected. There are two types of LVM sky fields. The first type are selected as regions showing extremely faint H$\alpha$ surface brightness ($<2.8\times10^{-18}$~erg~s$^{-1}$~cm$^{-2}$~arcsec$^{-2}$) in the Wisconsin H$\alpha$ Mapper (WHAM) \citep{haffner03}. We refer to these as the ``WHAM Darkest" sky fields. All these fields sit at high Galactic latitude ($|b|>30^{\circ}$), and are therefore often at high angular distances from their closest survey tiles. They provide the least contaminated measurements of sky emission possible, but they are not necessarily representative of the sky emission in the direction of the science fields. 

To compensate for this issue, we use a {\tt Healpix} tessellation \citep{gorski05} to divide the celestial sphere into a 768 element regular grid, and use a combination of low H$\alpha$ WHAM surface brightness, low stellar density in Gaia, and, when possible, the presence of nearby optically thick ($A_V>5$~mag) infrared dark clouds (IRDCs), to find the best possible sky field inside each element of the tessellation. We refer to these as ``Gridded" sky fields. They are subject to higher levels of MW contamination than the ``WHAM Darkest" fields, but since the grid elements have a mean separation of 7.3$^{\circ}$, they are ensured to be within a few degrees of the observed survey tiles \citep[see Figure 7 in][]{drory24}. 

For each science exposure, we randomly assign the SkyE and SkyW telescopes to a pair composed of the WHAM Darkest field and the Gridded field that are closest in angular distance to the survey tile that is being observed. In the extremely rare occasions in which no WHAM Darkest field is available, the closest Gridded field with $b>30^{\deg}$ is used instead. In Section \ref{sec:skysub} we discuss how the spectra in these two different fields is used to estimate the sky spectra towards the direction of the science tile being observed.

\subsection{Scheduling of Observations}
\label{sec:schedulling}

All the tiles identified within each of the LVM survey targets (see Section \ref{sec:tiles} and Table \ref{tab:targets}) are stored within a tile database. This database is accessed by our scheduler software, which provides the telescope with a feed of next tile positions in real time during the night. The same software package and database is used to track the successful completion of tiles as the survey progresses. Fields requiring dithered observations are still represented by a single tile in this framework, with the completion of individual dithers associated with the tile. There are a total of 28,928 tiles in our current database, representing $
\sim$10\% of the complete celestial sphere and $
\sim$13\% of the LCO observable sky at $\chi<2$. 

We do not expect to observe all of the tiles in the database during the 4.5 year baseline survey duration. Particularly, a fraction of the tiles belonging to the sparse high Galactic latitude grid, the Local Volume galaxies, and the Galactic plane extension are likely to remain unobserved at the end of the survey. However, the inclusion of the full area of these targets in the database is required to ensure that there are always suitable tiles available for observing throughout the course of the survey. The scheduler follows a set of conditions to choose the optimal tile for observation at any given time during the night. This takes into account the priority ranking of the target, the priority ranking of tiles within a given target, the airmass and Earth shadow height\footnote{The Earth shadow height is defined as the distance from the surface of the Earth to the location at which the observational line of sight intersects the Earth’s shadow. It is the primary parameter that defines the column of upper atmosphere gas that is directly exposed to solar radiation along the line of sight, and is therefore strongly correlated with the intensity of geocoronal emission \citep[e.g.][]{nossal01, gallant19}.} of the tiles at the time of the observation, and a set of lunar constraints (moon phase and lunar angular distance). 

The priority with which we observe our science targets is given in Table \ref{tab:targets}. The highest priority is always given to the LMC and the SMC. The next highest priority target is the core of Orion, with the Orion sparse grid having a slightly lower priority. Within the Milky Way plane, and at a lower priority than Orion, we prioritize (from highest to lowest) the main Galactic plane area, the Gum nebula, and then the extended plane. Below these core survey targets, we prioritize the Local Group Sample, followed by the 20 Mpc Sample, and finally the high Galactic latitude sparse grid, which is the lowest priority target. 

As many of the targets are composed of multiple tiles, we further allow for tile prioritization within each target. The LMC, SMC, Orion and Gum Nebula regions are prioritized inside out, such that the center is observed first. Individual tiles in the Galactic plane that overlap with the THOR HI survey area \citep{beuther16, wang20}, and tiles associated with known HII regions and SNRs are assigned a higher tile priority, so they are observed earlier in the survey.  

To observe a tile, we require an airmass $\chi<1.75$, except for the LMC for which we allow $\chi<2.0$ and M33 for which we allow $\chi<2.2$. To limit the contribution of geocoronal H$\alpha$ emission to our science spectra, a lower limit in Earth shadow height of 1000 km is employed for the MW plane and high Galactic latitude grid.  This is relaxed to 500 km for Orion and the Gum Nebula, where we expect to generally have a high surface brightness in our science target H$\alpha$ emission. No shadow height limits are imposed for extragalactic targets, including the Magellanic Clouds, as their redshift is sufficient to separate the atmospheric emission from the science target emission.

To meet our sensitivity requirements, we impose constraints on the moon brightness and angular distance, which also differ by target. Galactic tiles (including Orion, the Gum Nebulae, and the high Galactic latitude grid) can be observed under any lunar conditions, so long as an angular separation of 60$^\circ$ is met. Extragalactic tiles (including the LMC/SMC and Local Volume targets) are only observed with a maximum lunation of 0.25 (i.e. within 7.4 days of new moon) and a minimum angular lunar separation of 45$^\circ$. 

Tiles are selected based initially on their target and tile priorities, and then on their observability criteria. For dithered targets, to obtain uniform conditions over a subset of the dither positions, we only begin observing the tile if the observability conditions will be met for at least one set of three dithered observations. At the time the scheduler selects a given tile for observation, it also determines the optimal corresponding 12 spectrophotometric standard stars and two sky fields for simultaneous observation. Since the start of survey operations in late 2023, this scheduler software has successfully been operating to select tile and calibration sources throughout the night.

%% file: 5-data.tex
\section{Data Handling, Reduction, and Calibration} \label{sec:data}

\subsection{Data Flow Overview}

The raw LVM data are initially written and stored on site at LCO in a dedicated disk array capable of storing approximately 3 years of survey data. All LVM raw files are transferred in daily batches to the SDSS-V servers at the Center for High Performance Computing (CHPC) of the University of Utah. Once there, the data are processed through the LVM Data Reduction Pipeline \citep[DRP)][]{mejia25}, which ultimately produces a reduced and calibrated row-stacked spectra (RSS) file for each science exposure. 

The DRP output spectra of each science exposure are then processed through the LVM Data Analysis Pipeline \citep[DAP][]{sanchez25}. The DAP runs a custom spectral fitting algorithm on the individual spaxel spectra to model the stellar continuum and nebular emission lines. From these fits the DAP outputs best-fit emission line fluxes and kinematic parameters, stellar population parameters, and other derived quantities. The DAP outputs files for each science exposure are transferred back to the Utah CHPC SDSS-V servers.
 
LVM survey data products included in the SDSS-V internal product launches (IPLs) and public data releases (DRs) are made available through the SDSS Science Archive Server (SAS)\footnote{https://data.sdss5.org/sas/}. These products include the DRP output, containing reduced and calibrated RSS files for every science exposure, and the DAP output, containing best-fit model parameters including emission line parameters for each spaxel.

An interactive, browser-based visualization tool called the LVM Visualizer \citep[LVMvis,][]{katkov26_hips_LVMvis, SDSSCollaboration26b}\footnote{\url{https://dr20.sdss.org/lvmvis/}} provides access to LVM DRP and DAP products and helps users identify available observations and assess data quality before download or further analysis. The interface combines an Aladin Lite sky view \citep{baumann2022, aladinlite2024}, a table of LVM pointings, and an interactive spectral viewer in an adjustable layout. Through linked sky and table views, users can select pointings, apply column-based filters, inspect fiber locations, and view reduced spectra together with available best-fit models for individual fibers. The Aladin Lite component also allows LVM pointings to be displayed over HiPS maps from other surveys \citep{2015A&A...578A.114F}, with future releases planned to include HiPS maps based on LVM DAP parameters \citep{katkov26_hips_LVMvis}.

In this section we provide an overview of the data reduction and calibration process performed by the DRP and the spectral fitting performed by the DAP. A detailed description of the the LVM DRP is provided in \cite{mejia25}. A detailed description of the spectral fitting process performed by the DAP can be found in \cite{sanchez25}. 

\subsection{LVM Data Reduction Pipeline (DRP)}
\label{sec:drp}

\subsubsection{Detrending and Extraction}

For each exposure the LVM-I writes nine {\tt FITS} image files, one for each channel ($b,r,z$) in each of the three spectrographs. The DRP detrends the CCD images by applying pixel-to-pixel flats and masking bad pixels and cosmic rays. As discussed in Sections \ref{sec:obs-calib} and \ref{sec:stability}, we define epochs of instrument stability corresponding to date ranges within which a common set of calibration frames can be used to reduce the science data. For each epoch the DRP creates master calibrations files, including tracing and wavelength solutions from dome-flats and arcs respectively, and fiber-flats mapping the relative throughput of all fibers from the twilight-flats. After detrending, we extract the spectrum of each fiber in each channel using an efficient sampling matrix inversion algorithm which adequately recovers and deblends the signal in the cross-talk regions between neighboring fibers. 

After spectral extraction, the master wavelength solution of each camera is used to resample the spectra in each channel to a common wavelength grid with 0.5\AA\ pixels covering the 3600\AA$\le \lambda \le$9800\AA\ range. At this point, a fiber-flat correction that removes any fiber-to-fiber, telescope-to-telescope, and spectrograph-to-spectrograph variations in system throughput at fixed wavelength is applied. 


\subsubsection{Flux Calibration and Spectral Channel Combination}
\label{sec:fluxcal}

Flux calibration is performed for each channel ($b,r,z$) independently, prior to combining these channels into single spectra. The channel spectra are multiplied by a sensitivity curve to convert the signal from the instrumental units of electrons~s$^{-1}$~cm$^{-2}$~\AA$^{-1}$ to physical flux density units of erg~s$^{-1}$~cm$^{-2}$~\AA$^{-1}$. This step corrects for the instrumental sensitivity and removes the effects of atmospheric extinction. Notice that the DRP output spectra have units of total flux per spaxel. To convert the spectra to surface brightness units users must divide the fluxes and errors by the spaxel area (978.7~arcsec$^2$).

An independent sensitivity curve for each of the 10--12 standard stars observed during each exposure is calculated by taking the ratio of their LVM-I instrumental spectra, to a reference spectrum of their true brightness at the top of the atmosphere (taking into account airmass differences between the standards and the science field). A rough sky subtraction, sufficient for this purpose, is done on the standard star instrumental spectra prior to this calculation. The final sensitivity curve for each exposure is given by a weighted average of the sensitivity curves of the individual stars. The weights are given by the fraction of the total science exposure time during which each standard star was exposed.

The reference spectrum of each star comes from a best-fit stellar atmosphere model, which is obtained by performing a joint fit to the low resolution ($R\sim50$) Gaia BP/RP spectra, and the higher resolution continuum-nomalized LVM-I absorption line spectrum. The wealth of spectroscopic information on the relative strength of absorption features in the LVM-I spectra, combined with the overall shape of the optical continuum in the Gaia BP/RP spectra, provide good constraints on the stellar parameters of the F-type main sequence stars we select as standards. The averaging of 10--12 calibration sources helps us to average any systematic errors affecting the typing of individual stars.

In parallel to the procedure described above, the DRP calculates two other sets of alternative flux calibration sensitivity curves. The first uses the low resolution Gaia BP/RP spectra of the standard stars as the reference spectrum. The second also uses the Gaia BP/RP spectra, but the calculation is done on isolated bright stars that are well centered on spaxels in the science IFU. These are used mainly for validation and quality control purposes, as they are useful to evaluate the presence of systematic or catastrophic errors when running the standard calibration procedure. The different versions of the sensitivity curves are stored in the DRP output files and users can use them to rescale the spectra in order to change between them. An important caveat when using these curves is that the low resolution nature of the Gaia BP/RP spectra does not allow for good telluric absorption line corrections. 

Once the sensitivity curves are applied to the fiber spectra of each channel, the overlap regions between them share a common flux level and can be combined. Since the three channel spectra have been previously resampled to a common wavelength grid, no resampling is needed in this operation. We simply average the flux values of pixels in the overlapping spectral regions using inverse-variance weighting (with variances calculated as described in Section \ref{sec:errors}).

\subsubsection{Sky Subtraction}
\label{sec:skysub}

As discussed in Section \ref{sec:skyobs}, we obtain sky spectra at two different positions during each science exposure. The first is a high Galactic latitude ``WHAM Darkest" far sky field and the second is a ``Gridded'' dark near field within 10\degr of the science target. The sky we would like to remove has multiple components: an emission line component consisting of individually recognizable sky lines, a diffuse emission line component of many weaker lines from molecules that are not associated with the stronger lines, a continuum component, mainly due to moonlight and zodiacal light, and geocoronal emission, primarily from H$\alpha$ \cite[see, e.g.][]{noll25}. 

In the current version of the DRP, we model all of these components, except the stronger sky lines, as a pseudo-continuum. We extract this pseudo-continuum from the spectra  via an adaptive smoothing technique, which compares a smoothed version of a spectrum to the original spectrum and masks regions with positive excesses. We first construct a median spectrum from all of the fibers from each of the science, near sky and far sky fields. This is particularly important for the science spectrum, since some, but typically few, science fibers fall on bright stars, and the median filter restricts the contribution of these fibers to the spectrum we want to model. This is less of a problem for the sky spectra as the sky spectra positions were selected to exclude bright stars. 

We separate each median spectrum into continuum and line spectra. 
Having separate continuum and line contributions for each of the 3 spectra (science, near sky, and far sky fields), we assume that the sky spectrum in the science telescope is given by a scaled version of the continuum spectrum from the nearer ``Gridded" sky field summed with a scaled version of the line spectrum from the further (``WHAM Darkest") sky field. While better subtraction might be obtained by taking both the continuum and line contributions from the nearer sky field, these typically suffer from significant Galactic nebular emission contamination. The scalings are done by comparison of these components with the continuum and line components in the science field. 

We are aware that the simple approach described above is not ideal, and the testing and implementation of more sophisticated sky subtraction algorithms, using a variety of radiative transfer-based model calculations, such as the ESO Sky Model \citep{noll12}, are being actively developed, but these are not incorporated into the current version (v1.2.0) of the LVM DRP. However, even though our current approach is fairly basic, the error in subtraction for most lines is less than 5\% of the line flux.

Geocoronal emission is treated separately by calculating global nightly fits to the observed variation of the H$\alpha$ line flux in the ``WHAM Darkest" fields, as a function of Earth shadow height. Our estimate of the geocoronal H$\alpha$ emission is not currently subtracted from the spectra, but is recorded (in the headers) of the data products, so it can be incorporated into the analysis. As mentioned above, sky subtraction and geocoronal emission modeling is an active area of development within the LVM collaboration, and we expect to present an update on our sky subtraction strategy in a future publication.

\subsubsection{Masking and Error Propagation}
\label{sec:errors}

Throughout all the processing steps, the DRP tracks bad detector pixels and pixels affected by cosmic rays through a pixel mask extension that propagates through the final data products. Standard error propagation is also performed through all the steps of the pipeline. When calculating errors, we account for read-noise and assume Poisson errors for photo-electron counts on the detector. We also treat all pixels in the spectra as independent measurements. The latter is formally correct in all steps prior to wavelength resampling, but might translate into a small level of error underestimation for operations done after this step. This is because of the covariance introduced between neighboring pixels when interpolating the spectra into a common wavelength grid. 


\subsection{LVM Data Analysis Pipeline (DAP)}

The LVM DAP is presented and thoroughly described in \cite{sanchez25}. The pipeline performs spectral fits to the individual spaxel spectra in the DRP output. The fitted model includes nebular emission lines, a stellar continuum component, and dust extinction. The pipeline outputs emission line parameters (flux, centroids, and widths) for 192 transitions \citep[see Appendix A in][]{sanchez25}, and the template coefficients describing the best-fit stellar continuum model, which can be translated into probability-distribution-functions (PDFs) for the stellar parameters (effective temperature, surface gravity, and metallicity) of the stars that populate each spaxel.

The LVM DAP differs from standard spectral fitting algorithms typically used in IFU surveys of galaxies in the adoption of a resolved stellar population (RSP) synthesis approach for modeling the stellar continuum. In the LVM data, the resolution is such that neither single stars, nor fully sampled stellar populations, populate the individual spaxels. Therefore, the continuum cannot be fitted using neither individual stellar templates, nor stellar population synthesis models that assume a fully sampled IMF, integrated using evolutionary tracks over a particular star formation history. Instead, the continuum is modeled through a linear combination of templates, which are built by applying a non-supervised clustering method to the MaSTAR stellar library \citep{yan19}. Each template is representative of sub-samples of stars with similar SEDs (i.e. one cluster\footnote{Here cluster refers to a group of stars identified by the clustering algorithm and not to a stellar cluster as a physical structure.}), and is associated with the PDF of best-fit stellar parameters of the stars in each cluster, measured using the {\tt CoSHa} algorithm \citep{mejia22}. The RSP approach provides the adequate flexibility to produce a good description of the stellar spectra seen across the LVM data, allowing the proper separation of stellar continuum and nebular emission lines.

The DAP is run automatically on all the DRP output files. The means that each DAP output file is associated with a single DRP output file corresponding to an individual LVM-I exposure. This is suitable for the MW survey where we obtain a single exposure per tile. For tiles in the Magellanic Clouds and the local volume galaxies, the DAP output for each individual dithered exposure does not exploit the increase in depth associated with combining the nine dithers. The latter requires resampling and combining the dithered spectra onto a regularly sub-sampled data-cube, and then running the DAP (or other spectral fitting routine) on the individual data-cube spaxels. In the future we plan to implemented this as part of the regular LVM survey data processing workflow. A general use data-cube generation tool called {\tt 3DCubeGen}, which is optimal for processing the LVM data, has been recently published by \cite{ibarra-medel26}\footnote{https://github.com/hjibarram/3DCubeGen}

%% file: 6-products.tex
\section{LVM Data Products} \label{sec:products}

In this section we describe the LVM DRP and DAP output files that serve as the main LVM data products. We focus only on the most important products, which most LVM users are likely to conduct scientific analysis on. Both pipelines produce a series of other intermediate, ancillary, and quality assessment products that are not described here. We refer the reader to the DRP \cite{mejia25} and DAP \cite{sanchez25} articles, and to the public SDSS data releases documentation for further details.

\subsection{Main DRP Output Files}

The main output of the DRP is the {\tt lvmSFrame-{EXPNUM}.fits} file. Where {\tt EXPNUM} refers to the unique ID of an individual LVM-I exposure. This is a single multi-extension {\tt FITS} file that contains the following extensions:

\begin{itemize}
    \item {\tt "PRIMARY"}: primary header containing exposure metadata. This includes  a series of instrumental, targeting, observational, and data reduction parameters, as well as ephemerides, sky conditions, and quality assessment information. Image quality and astrometric information derived from the co-added guider images is also stored here.
    
    \item {\tt "FLUX"}: calibrated and sky subtracted spectra for all LVM-I spaxels (including the Sci, SkyE, and SkyW IFUs, and the individual Spec telescope fibers). The spectra is stored as a 2D {\tt FITS} image in row-stacked-spectrum (RSS) format, with flux units of erg~s$^{-1}$~cm$^{-2}$~\AA$^{-1}$ per spaxel. 

    \item {\tt "IVAR"}: inverse variance of the {\tt FLUX} spectra. The error in the {\tt FLUX} spectra is given by $\sqrt(1/{\tt IVAR})$. The spectra is stored as a 2D {\tt FITS} image in RSS format, with flux units of (erg~s$^{-1}$~cm$^{-2}$~\AA$^{-1}$)$^{-2}$ per spaxel. 

     \item {\tt "MASK"}: pixel mask spectra. The mask is stored as a 2D {\tt FITS} image in RSS format, and contains pixels masked due to detector defects, cosmic rays, and saturation.
 
     \item {\tt "WAVE"}: wavelength array corresponding to all spectra in the other RSS extensions. The array is stored as a 1D {\tt FITS} image, and contains air wavelength in units of Angstroms.

     \item {\tt "LSF"}: line-spread-function FWHM spectrum for all LVM-I spaxels. The wavelength dependent LSF width for each spaxel, corresponds to a 4th order polynomial fit to the measured Gaussian FWHMs (in \AA units) of a series of arc-lines across the three channels. The LSF FWHM is stored as a 2D {\tt FITS} image in RSS format, and has units of Angstroms.

     \item {\tt "SKY"}: sky spectra for all LVM-I spaxels, which was subtracted for creating the {\tt FLUX} spectra. The sky spectra is stored as a 2D {\tt FITS} image in row-stacked-spectrum (RSS) format, with flux units of erg~s$^{-1}$~cm$^{-2}$~\AA$^{-1}$ per spaxel.

     \item {\tt "SKY\_IVAR"}: inverse variance of the {\tt SKY} spectra. The error in the {\tt SKY} spectra is given by $\sqrt(1/{\tt SKY\_IVAR})$. The spectra is stored as a 2D {\tt FITS} image in RSS format, with flux units of (erg~s$^{-1}$~cm$^{-2}$~\AA$^{-1}$)$^{-2}$ per spaxel. 

     \item {\tt "SLITMAP"}: binary {\tt FITS} table containing the parameters associated with each spaxel. Information like the unique ID number of each fiber, the telesope and spectrograph associated with it, its position in the focal plane IFU and along the spectrograph slit-head, and flags associated with the health of the fibers, are all reported here. Most importantly, the equatorial coordinates of each spaxel, calculated from the AG cameras based astrometric solution are provided here. 
     
\end{itemize}

Besides the main output file described above, the DRP output also includes files, in similar format, containing the spectra prior to sky subtraction ({\tt lvmCFrame-{EXPNUM}.fits}), and prior to the combination of the individual spectral channels (b,r,z), both before and after the wavelength resampling, flat-fielding, and flux calibration steps ({\tt lvmFrame-{CHANNEL}-{EXPNUM}.fits} and {\tt lvmFFrame-{CHANNEL}-{EXPNUM}.fits} respectively).

\subsubsection{Main DAP Output Files}

The DAP output files are thoroughly described in Section 3.4 \cite{sanchez25} (see their Tables 1 and 2). The main DAP data products are the 
{\tt dap-rsp108-sn20-{EXPNUM}.dap.fits} files, which contain the results of the spectral fitting procedure which is run for the spectra of all spaxels in each {\tt lvmSFrame} file produced by the DRP.

Readers should refer to \cite{sanchez25} for a description of the different fit parameters. In brief, these include emission line parameters (flux, centroids, and line-widths) from parametric Gaussian fits for 15 strong emission lines, and from moment-based non-parametric measurements for a full set of 191 transitions (see Table 8 in \citealt{sanchez25}), continuum line-of-sight stellar velocity distribution (LOSVD) first and second order parameters (velocity, and velocity dispersion), and the best-fit coefficients for the base of templates used in the stellar continuum fitting.

%% file: 7-data_quality.tex
\section{Instrument Performance, Data Quality, and Survey Progress} \label{sec:data_quality}

\begin{figure*}[th]
\gridline{\fig{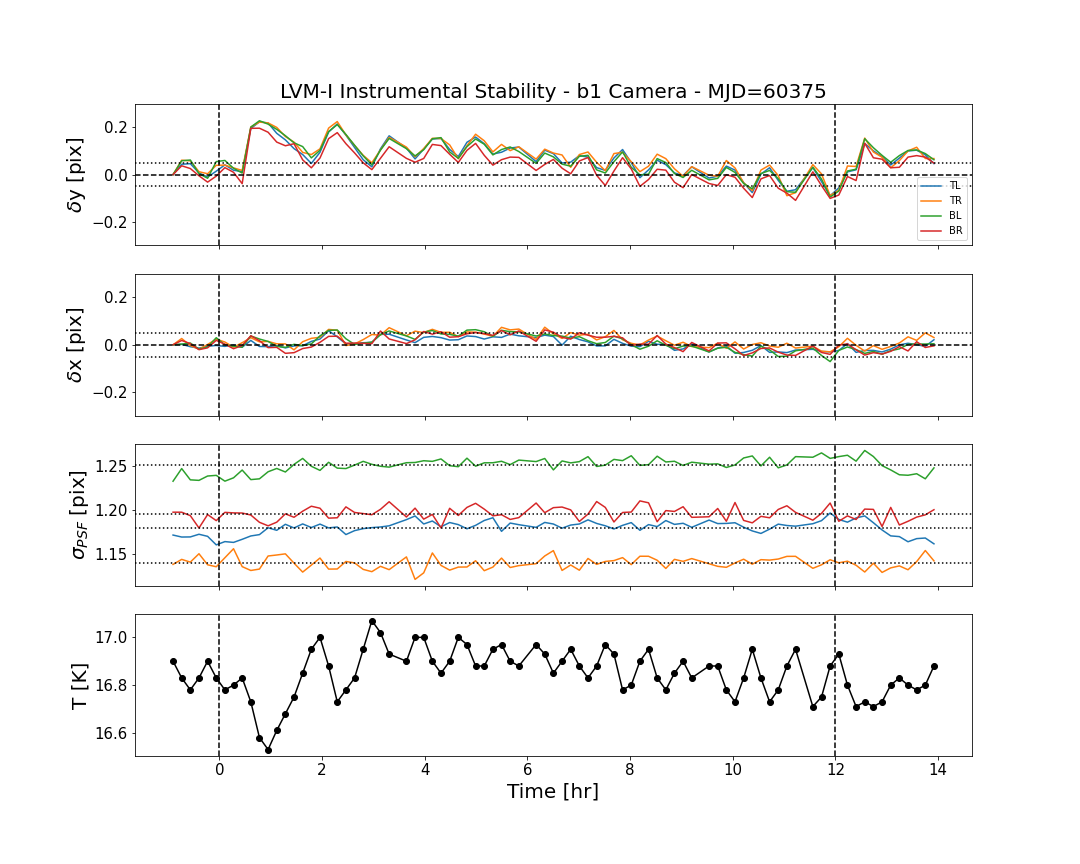}{1.0\textwidth}{}}
\vspace{-1.5\baselineskip}
\caption{Time series of PSF vertical (1st panel) and horizontal (2nd panel) displacements, PSF width (3rd panel), and cryostat window temperature, over a full day, for the arc lines closest to the four corners of the LVM-I Spec-1 $b$ channel detector (Top-Left, Top-Right, Bottom-Left, Bottom-Right). All other channels in all spectrographs present similar behaviors. In the top two panels the dotted horizontal lines mark displacements of 0.05 pixels. In the third panel the horizontal dotted lines mark the average PSF width of each arc line. In all panels the vertical dashed lines mark the times of the liquid nitrogen (LN2) fills, which induce large temperature swings (these happen twice per day, outside observing hours). 
}
\vspace{0.5\baselineskip}
\label{fig:stability}
\end{figure*}

\begin{figure*}[t]
\gridline{\fig{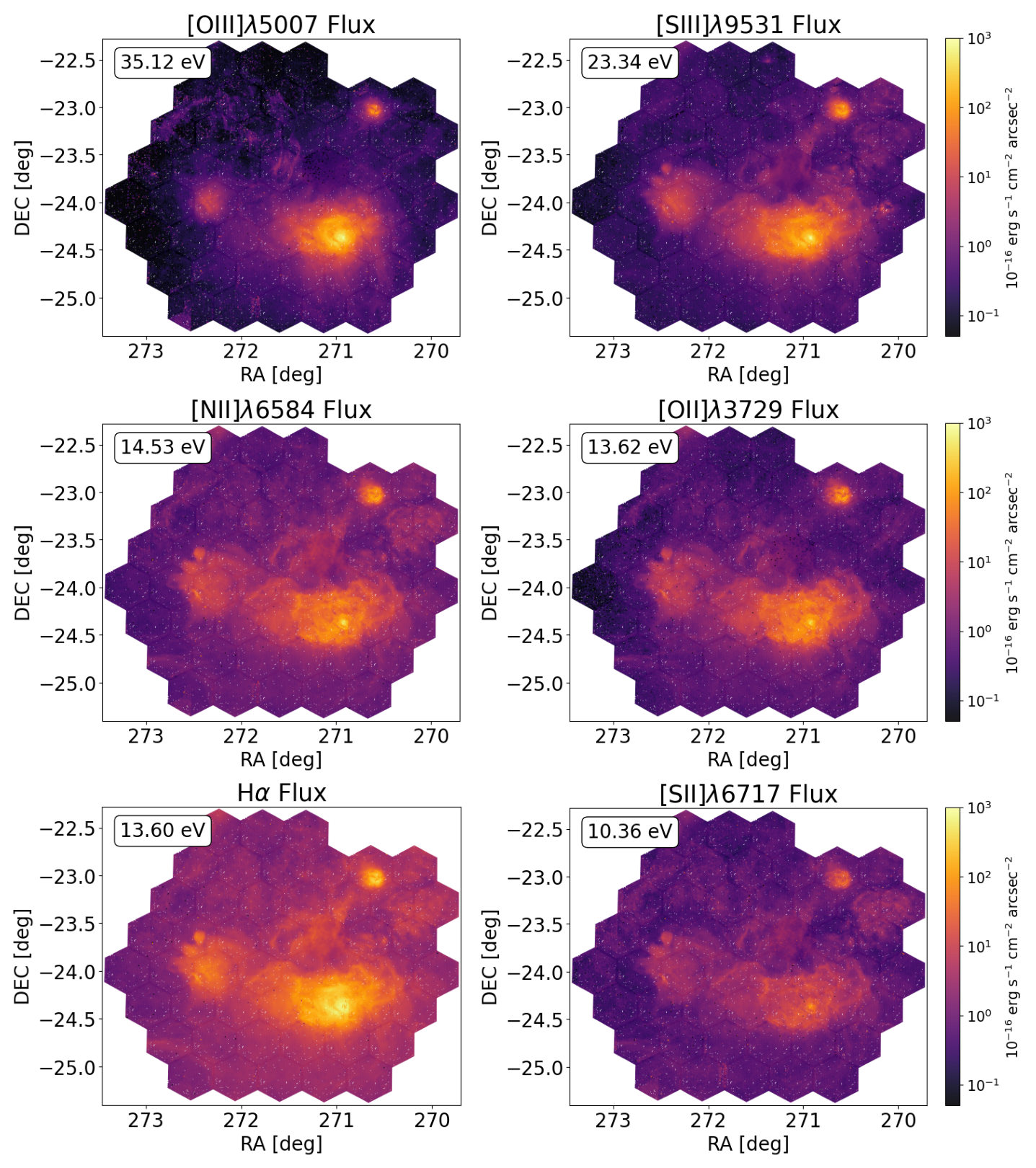}{0.95\textwidth}{}}
\vspace{-1.5\baselineskip}
\caption{Emission line flux maps of an approximately  $4^{\circ} \times 3^{\circ} $ region covering the Lagoon nebula (M8) towards the SW of the center of the image, the Trifid nebula (M20) towards the NW, the Siemis 188 complex (about $1^{\circ}$ East of M8), and the W28 supernova remnant (just SW of M20). Maps are presented for the H$\alpha$ recombination line, and five commonly used strong collisionally excited nebular lines ([OIII]$\lambda$5007, [SIII]$\lambda$9531, [NII]$\lambda$6584, [OII]$\lambda$3729, and [SII]$\lambda$6717). The panels are sorted by decreasing ionization potential for the emitting ions (labeled in each panel). At every pixel in these images LVM delivers a deep medium-resolution optical spectrum.
}
\vspace{0.5\baselineskip}
\label{fig:lagoon}
\end{figure*}

\begin{figure*}[t]
\gridline{\fig{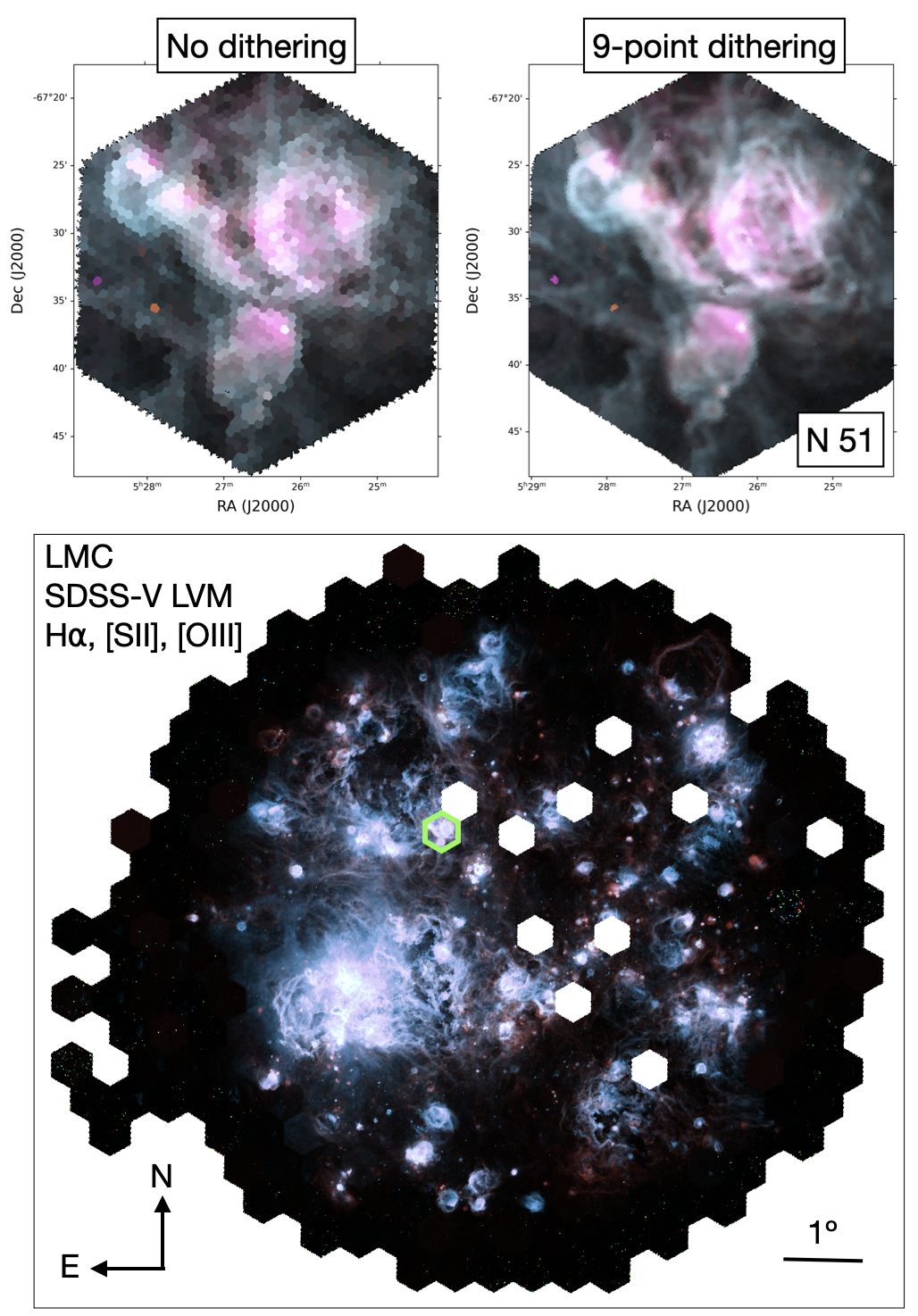}{0.75\textwidth}{}}
\vspace{-1.5\baselineskip}
\caption{{\it Bottom panel:} SDSS-V LVM H$\alpha$;[SII]$\lambda$6717;[OIII]$\lambda$5007 emission color composite map of the LMC, based on currently available data of the galaxy. The tile covering the N51 HII region is highlighted in green. {\it Top panel:} Nebular emission color-composite reconstructed image of the N51 HII region based on a single exposure with no dithering (left) and using the 9-point dithering scheme adopted in LVM for observations of the Magellanic Clouds and other local volume galaxies (right). 
}
\vspace{0.5\baselineskip}
\label{fig:lmc}
\end{figure*}

\subsection{Instrumental Stability}
\label{sec:stability}

A core assumption behind the LVM calibration and data reduction strategy is that the instrument is stable enough to allow the use of calibration frames not taken back to back with the science data. Calibration frames are used for tracing the position of the fiber spectra on the detector and for calculating wavelength solutions. In many fiber-fed IFUs (e.g. PPAK and MaNGA) instrumental flexure, fiber-torsion, and thermal effects translate into systematic errors in the extracted flux, wavelength calibration, and line-spread-function (LSF) of the spectra. This requires calibrations to be taken back-to-back with every science observation, which has a significant impact in terms of survey efficiency. 

To avoid these effects and achieve the required level of stability, LVM-I is designed to have a static fiber system that feeds gravity-invariant spectrographs hosted in a thermally-controlled environment. The goal of this design is to ensure that any movement of the fiber images on the detectors consists of small ($< 0.5$ pixel) simple translations, with negligible changes in the higher order ($\mathcal{O}\ge1$) terms of the tracing and wavelength solution polynomials. This allows the use of calibration frames taken hours, even days, apart from the science data by applying simple offsets to the zeroth order terms of these polynomials. These offsets can be easily measured from the science data itself by measuring the wavelengths of known sky lines and by cross-correlating the spatial profile of the fibers in the science and calibration frames. In terms of LSF stability we worked towards a goal of keeping variations in the width of the fiber PSF below the 5\% level.

During commissioning we conducted an experiment to evaluate the instrumental stability of the LVM-I in order to validate the above assumptions. For this test we obtained a sequence of arcs with a 10 minute cadence over a continuous 15 hr period. For each of the nine cameras we measured the detector coordinates of four non-blended arc lines near the corners of the chip. We do this by fitting two dimensional Gaussians to these four monochromatic fiber images, which also yields a measurement of the width of the PSF ($\sigma_{PSF}$). The sequence covers the morning and evening liquid nitrogen fills of the cameras, and therefore samples the full range of temperature variations suffered by the spectrographs on a daily basis.

Figure \ref{fig:stability} presents the results of this experiment for the blue channel camera of spectrograph 1. We observe a very similar behavior in all nine LVM-I cameras both in qualitative and quantitative terms. The first and second panels show the cross-dispersion (y) and dispersion (x) direction centroid offsets of the arc lines in the four corners of the chip (labeled top-left, top-right, bottom-left, and bottom-right). The third panel shows the circularized width ($\sigma_{PSF}=\sqrt(\sigma_{x}*\sigma_{y})$) of the arc line images. The bottom panel shows the temperature in front of the cryostat window as a function of time. Besides the 0.2$^{\circ}$-0.3$^{\circ}$~C temperature swings associated with the LN2 fills, over a night the instrument is thermally stable at the $\sim 0.1^{\circ}$~C level. 

We observe almost negligible offsets ($<0.05$ pixels) in the dispersion direction. In the cross-dispersion direction on the other hand, we observe small offsets of up to 0.2-0.3 pixels that are highly correlated with the instrumental temperature \footnote{We observe an anticorrelation between the dispersion direction offsets of the $b,z$ channels and those of the $r$ channel. Given the double-dichroic design of the spectrographs this implies that the thermal effects behind these offsets happen in the vertical direction either at the collimator mirror assembly or at the pseudo-slit assembly.}. These thermally-induced offsets are sufficiently small as to not affect the higher order terms of the tracing and wavelength solution polynomials, and can therefore be trivially corrected in the pipeline. The circularized width of the monochromatic fiber images shows $<2$\% variation over the whole period. These results confirm that the LVM-I instrumental stability meets the requirements of the LVM survey, and validates the adopted calibration strategy.

\subsection{Overview of Early LVM Data}

Here we showcase spectral line maps of two HII regions in the MW and the LMC respectively, to provide an overview of the quality of the data in these two main target distance and resolution regimes. The maps presented below are built directly from the line fluxes available in the DAP output files without applying any scalings or other corrections. 


In the MW we showcase a roughly $4^{\circ} \times 3^{\circ} $ region covering the Lagoon nebula (M8) towards the SW \citep{Singh26}, the Trifid nebula (M20) towards the NW \citep{sattler26}, the Siemis 188 complex (about $1^{\circ}$ East of M8), and the W28 supernova remnant (just SW of M20). Figure \ref{fig:lagoon} presents maps of strong collisionally excited lines (CEL) of oxygen, sulfur, and nitrogen ions with different ionization potentials, together with the H$\alpha$ hydrogen recombination line. These are built from a mosaic of 49 LVM-I science IFU tiles. The CEL maps are sorted from top-left to bottom-right in order of decreasing ionization potential for the corresponding ions. 

They are constructed by directly plotting the DAP line fluxes derived from single Gaussian profile fits to the emission lines in approximately 86,000 spectra. The amount of data presented in Figure \ref{fig:lagoon} corresponds to less than two nights of LVM survey operations, which highlights the high mapping efficiency that the ultra-wide-field LVM-I IFU provides.

The ionization structure of the different nebulae can be clearly traced in the LVM data. In the HII regions, high ionization species like O$^{++}$ trace the inner regions closer to the main ionization sources. Intermediate ionization species like S$^{++}$ emit through most of the volume of the nebulae. And low ionization species like O$^+$, N$^+$, and S$^+$ trace the outer regions of the HII regions, where the transition to the diffuse ionized medium occurs, and are particularly bright in the W28 supernova remnant. 


These data show that the LVM can resolve the inner structure of Galactic ionized nebulae at the sub-pc to pc level. The wide spectral range and depth of the spectra permit the detection of H and He recombination lines, as well as forbidden and auroral metal lines used to diagnose the physical conditions of the gas \citep[e.g. electron density, temperature, ionization parameter, and chemical abundances][]{sattler26}. In the brightest regions of these nebulae even ultra-faint metal recombination lines, three to four orders of magnitude fainter than H$\alpha$ can be detected \citep{Singh26}. The integral-field nature of the survey permits the mapping of complete structures from the location of the ionizing sources out to the transition between discrete nebular and the diffuse medium.

\begin{figure*}[th]
\gridline{\fig{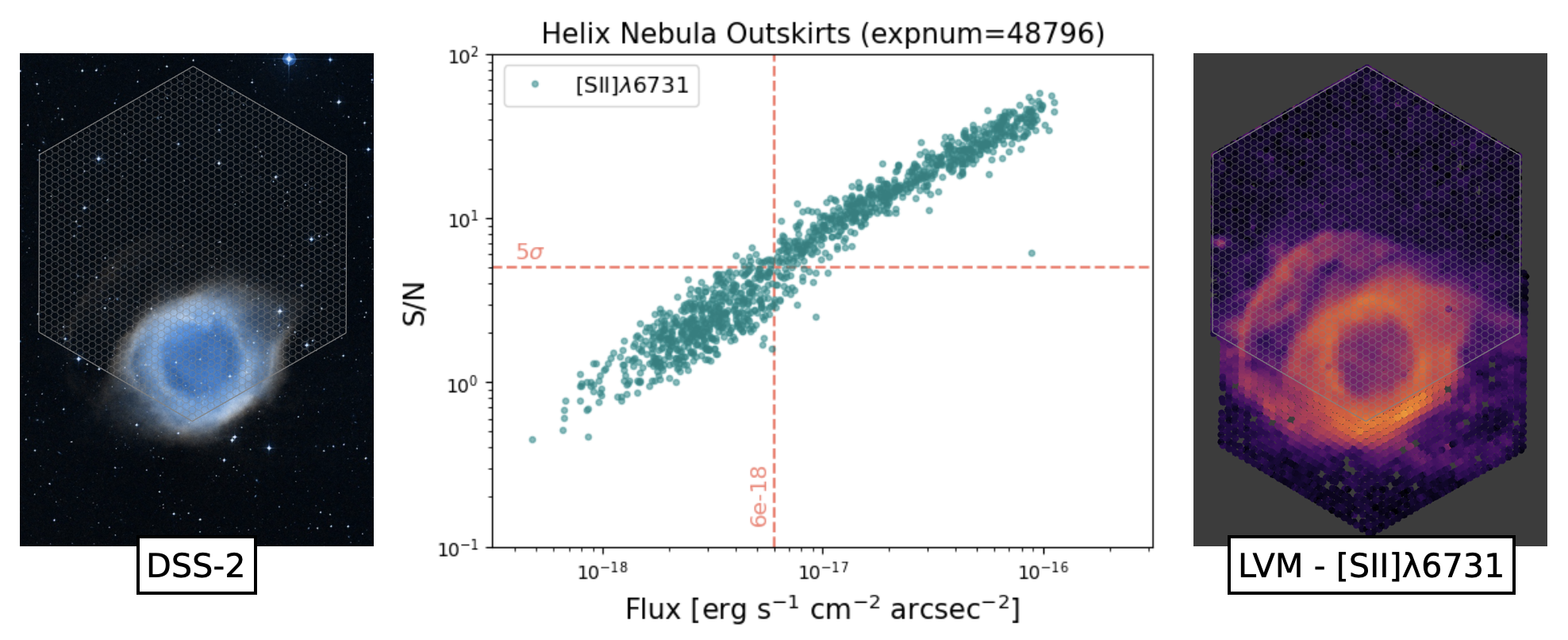}{0.95\textwidth}{}}
\vspace{-1.5\baselineskip}
\caption{Signal-to-noise (S/N) as a function of [SII]$\lambda$6731 emission line flux for all spaxels in a 900~s exposure of the outskirts of the Helix planetary nebula (central panel). The horizontal and vertical dashed red lines mark the LVM program science requirement of reaching a 5$\sigma$ line depth of $6\times 10^{-18}$~erg~s$^{-1}$ cm$^{-2}$~arcsec$^{-2}$ for the single epoch 900¬s exposures used in the MW survey. The left and right panels shows a color-composite broad-band image of the nebula \citep[from the DSS2 survey][]{lasker96} and the SDSS-V LVM [SII]$\lambda$6731 emission line map respectively. The tile corresponding to the data in the central panel is highlighted in grey.
}
\vspace{0.5\baselineskip}
\label{fig:depth}
\end{figure*}

For the LMC, Figure \ref{fig:lmc} presents the H$\alpha$ map of the complete coverage we have obtained to date on the galaxy. The figure also zooms into the the LHA120-N51 star forming region \citep{henize56}. This is a complex nebular structure composed of multiple HII regions ionized by recent massive star formation, which also encompasses DEM L205, a 35,000 year old supernova remnant confirmed by X-ray and radio emission, sitting on the NE end of the complex \citep{maggi12}. 

The zoom-in maps on LHA120-N51 are color-composites of the of H$\alpha$ (orange), [SII]$\lambda \lambda$6717,6731 (blue), and [OIII]$\lambda$5007 (purple) emission lines. Two versions of the maps are presented. The top left panel shows a map constructed from a single 900~s exposure (i.e. with no dithering), equivalent to the observing strategy that the LVM adopts in the MW. At the distance of the LMC \citep[$49.59\pm0.54$~kpc][]{pietrzy19} the LVM-I IFU spaxel diameter corresponds to a physical scale of 8.49~pc. It can be seen that at these scales the sampling of the IFU spaxels hampers our ability to resolve small structures like bubbles, filaments, and dust-lanes.

The top-right panel of Figure \ref{fig:lmc} shows the same map, but constructed from nine dithered exposures following the LVM observing strategy for extra-galactic targets (Section \ref{sec:obs-science} and Figure \ref{fig:dither}). The image reconstruction is performed using {\tt 3DCubeGen} \citep{ibarra-medel26}. The 9-point dithering strategy largely overcomes the effects of spaxel sampling and significantly increases the fidelity of the reconstructed line maps. The multiple exposures are also needed to reach the stellar continuum depth required by stellar population studies in these external galaxies \cite[see science requirements discussion in][]{drory24}.

The DEM L205 supernova remnant can be clearly distinguished at the NE end of the region. It appears as the blue colored bubble, with a well-defined shock front with a high [SII]/H$\alpha$ ratio, characteristic of these types of objects. This also demonstrates the power of the LVM spectral information for characterizing the nature of the physical processes that set the energetics and ionization state of the ISM.

\subsection{Line Flux Depth and Stacking Limits}

\cite{drory24} presents the LVM science requirements, which include a need to achieve an H$\alpha$ 5$\sigma$ emission line flux depth of $6\times 10^{-18}$~erg~s$^{-1}$ cm$^{-2}$~arcsec$^{-2}$ in 900~s exposures taken in dark time. In order to evaluate whether the survey reaches the required depths, we consider a plot of the [SII]$\lambda$6731 emission line flux vs signal-to-noise for a single LVM survey tile which maps the outskirts of the Helix planetary nebula in Figure \ref{fig:depth}. This exposure was taken in dark time, and under photometric conditions.

We choose to use the [SII]$\lambda$6731 line instead of H$\alpha$ in this test because it very close in wavelength to the latter (such that the sensitivity of the instrument is basically the same as for the wavelength at which the requirement is defined) but does not suffer from geocoronal contamination, which can be significant for H$\alpha$ in low surface brightness regions. Figure \ref{fig:depth} clearly shows that the LVM-I meets its design specifications in terms of sensitivity.

Another interesting aspect of the LVM data is the possibility of binning neighboring spaxels to increase the depth of the spectra, at the expense of spatial resolution. This is particularly relevant for the study of low surface brightness diffuse ionized structures, like those associated with MW extra-planar ionized gas at high Galactic latitudes. The LVM survey was designed to be able to reach down to 5$\sigma$ emission line surface brightness as low as $1\times 10^{-18}$~erg~s$^{-1}$ cm$^{-2}$~arcsec$^{-2}$ when stacking of the order of 100 neighboring spaxels together. 

To test the limits down to which line depth can be pushed in LVM via stacking, we conduct a simple experiment using a high Galactic latitude ``sparse grid" science tile. The chosen field is close to the Galactic South pole, and was observed during dark time and under photometric conditions. This tile is practically void of signal, making it an ideal target for conducting this evaluation, as there are no measurable astrophysical surface brightness fluctuations within the tile, which could bias the spectral noise statistics when stacking the spectra of neighboring spaxels. 

\begin{figure}[t]
\gridline{\fig{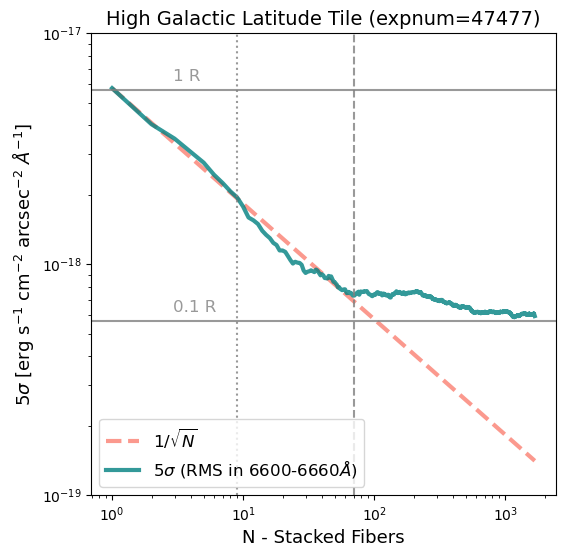}{0.45\textwidth}{}}
\vspace{-1.5\baselineskip}
\caption{Decrease in spectrum noise as a function the number of stacked fibers. The noise (teal solid line) is measured as the flux r.m.s. in clean wavelength window (6600-6660~\AA), and the data corresponds to the IFU spectra of a high Galactic latitude tile which is practically devoid of astrophysical signal. The dashed salmon line shows the 1/$\sqrt{N}$ behavior expected in the absence of systematic error. The vertical dotted line marks 9 fibers (equivalent to the number of dithers LVM uses for extragalactic targets), and the vertical dashed line marks the limit of 70 fibers, above which systematic errors in the spectrum flux start to dominate, limiting the usefulness of stacking to push the surface brightness sensitivity of the data. The two horizontal grey lines mark the H$\alpha$ surface brightnesses corresponding to 1~R and 0.1~R
}
\vspace{0.5\baselineskip}
\label{fig:stacking}
\end{figure}

Figure \ref{fig:stacking} shows the result of a simple experiment, in which we measure the pixel-to-pixel flux r.m.s. in a clean spectral region (6600\AA--6660\AA), free of significant nebular emission and sky subtraction residuals. This empirical noise estimate was measured for an individual science fiber, and then for a series of stacked spectra built by averaging the spectra of anywhere from 2 to over 1,000 neighboring fibers. Figure \ref{fig:stacking} presents the per pixel 5$\sigma$ empirically measured noise level in the spectra as a function of the number of stacked fibers.

The depth of the spectra increases following a standard $1/\sqrt{N}$ statistical behavior in the $N<70$ regime. Beyond that limit, the curve flattens towards a systematic error floor of $\sim 6\times 10^{-19}$~erg~s$^{-1}$ cm$^{-2}$~arcsec$^{-2}$ (approximately 0.1~R)\footnote{The Rayleigh (R) is a unit of flux ($F$) defined as the integral over all solid angles (4$\pi$) of an intensity (i.e.\ a surface brightness) $\mu=10^6$\,photons\,cm$^{-2}$\,sr$^{-1}$\,s$^{-1}$. So $F$\,[R]=4$\pi\times10^{-6}\mu$\,[photons\,cm$^{-2}$\,s$^{-1}$]. For H$\alpha$ photons with energy $h\nu_{\rm H\alpha}$, 1\,R corresponds to a surface brightness $\mu_{1R, H\alpha}=5.67\times10^{-18}$\,erg\,s$^{-1}$\,cm$^{-2}$\,arcsec$^{-2}$.}, which cannot be crossed regardless of the number coadded. This limit probably reflects a floor in systematic error in the extraction, flat-fielding, and calibration of the LVM spectra. The desired goal of reaching surface brightness limits of the order of $1\times 10^{-18}$~erg~s$^{-1}$ cm$^{-2}$~arcsec$^{-2}$ is reached after stacking 20-30 neighboring fibers, and is therefore within the reach of the LVM data.

\subsection{Spectral Resolution and Line Spread Function}

\begin{figure}[th]
\gridline{\fig{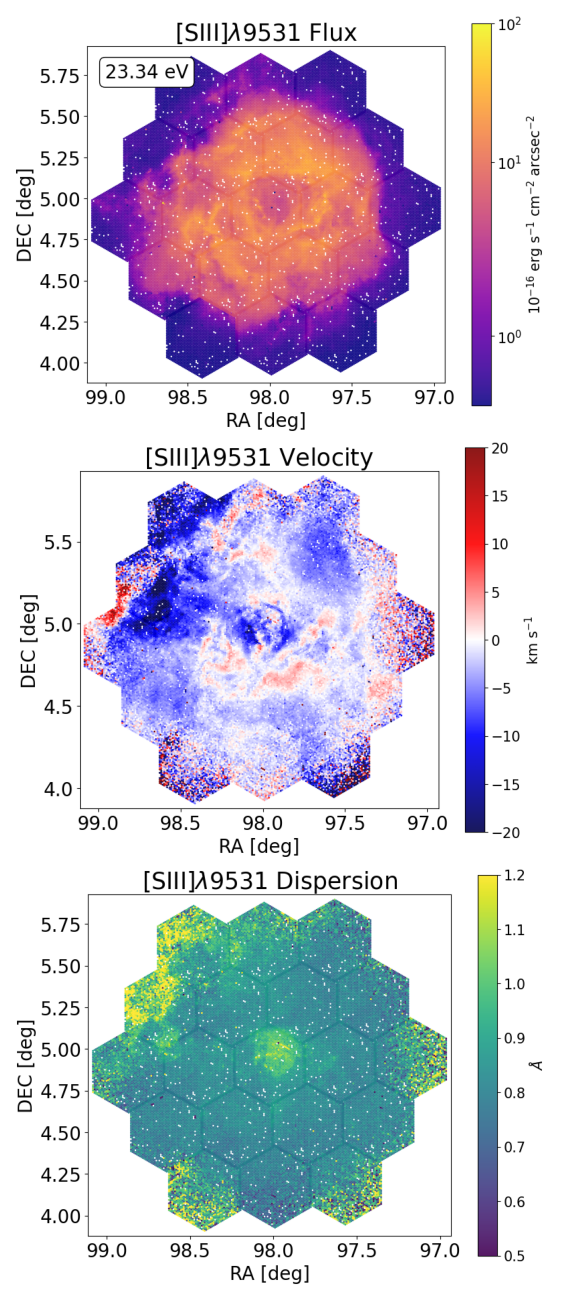}{0.47\textwidth}{}}
\vspace{-1.5\baselineskip}
\caption{The Rossette Nebula (NGC2237) as seen by LVM. The top panel shows the intensity of the [SIII]$\lambda$9531 emission line. The middle and bottom panels present maps of the best-fit heliocentric line of sight velocity of line, and the the best-fit line width ($\sigma$ in \AA) from the DAP Gaussian fits.
}
\vspace{0.5\baselineskip}
\label{fig:kinematics}
\end{figure}

The LVM spectra LSF is quantified by the DRP by measuring the Gaussian width of a series of isolated emission lines in the arc-lamp spectra, and fitting a low order polynomial to the best-fit Gaussian line-widths as a function of wavelength. The bottom panel of Figure \ref{fig:channels} presents these measurements for the three spectral channels. The average instrumental line-spread-function width ($\sigma_{LSF}$) of each channel covers a range of 36-59~km~s$^{-1}$ in the $b$-channel, 23-31~km~s$^{-1}$ in the $r$-channel, and 24-32~km~s$^{-1}$ in the $z$-channel (see also Section \ref{sec:spectrographs}). These curves, which are measured from the master-arcs used to do the wavelength calibration, are provided in the DRP output files for each science exposure. 

Figure \ref{fig:kinematics} shows maps of line flux and line kinematics (velocity and dispersion) of the Rosette nebula, derived from the [SIII]$\lambda 9531$ transition. The high atomic mass of sulfur, combined with the brightness of this strong nebular line, and the long wavelength of this transition (which minimizes the effects of dust extinction and maximizes the physical velocity resolution of the spectra), make it an excellent kinematic tracer in the LVM data. Bulk ionized gas motions at the level of a few km~s$^{-1}$ can be easily detected across this nebula in the LVM data. In the linewidth map, a high gas velocity dispersion ($\sim$30 km~s$^{-1}$) structure in the center of the nebula, which coincidences with the extent of the X-ray emitting bubble, and is most likely associated with stellar-wind induced shocks, is also easily noticeable. See \cite{villa-durango25} for a more detailed kinematic analysis of the LVM data on the Rosetta.

The third panel of Figure \ref{fig:stability} shows that the LSF is stable to thermal variations at the $\sim$0.01 pixel (i.e. $\lesssim 0.5$ km~s$^{-1}$). Nevertheless, the native LSF is of course different for different fibers, and can change whenever the spectrographs are refocused after servicing activities during the survey. 

It is important to highlight that while the LVM-I LSF is well described to zero order by a Gaussian function, deviations from Gaussianity (mainly in the form of kurtosis and the presence of asymmetric wings) are present, especially at the low and high wavelength ranges of each channel, and are stronger for fibers that are imaged on the top and bottom parts of the detectors. A more detailed characterization of the LVM data LSF has not been performed by the LVM team so far, and should be done by users in order to do high-precession kinematic analysis of the data.

\begin{figure}[t]
\gridline{\fig{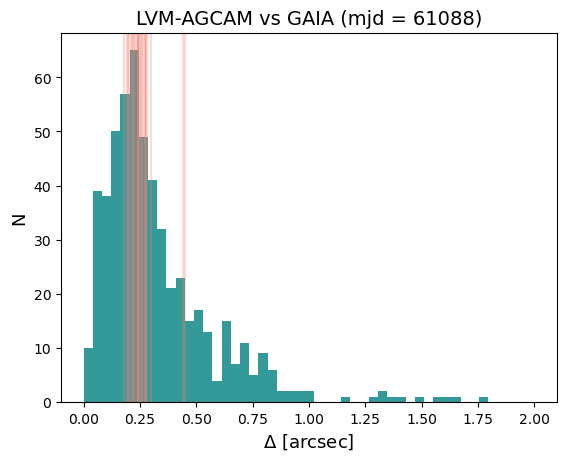}{0.45\textwidth}{}}
\vspace{-1.5\baselineskip}
\caption{Distribution of absolute offsets ($\sqrt{\delta x^2+ \delta y^2}$) for GAIA stars in all Sci telescope AG camera images taken during the night of mjd=61088, associated with the uncertainty in the AC camera astrometric solution. Vertical lines mark the average residual offset for each analyzed image.)
}
\vspace{0.5\baselineskip}
\label{fig:astrometry}
\end{figure}

\subsection{Astrometric Precision}
\label{sec:astrometry}

The astrometric solution of the science IFU is derived from the coadded AG camera images described in Section \ref{sec:obs-science}. A complete plate-scale solution of the focal plane using the two blanking AG camera coadded images is performed using {\tt astrometry.net}. This solution is then propagated to the science IFU to calculate the equatorial coordinates corresponding to the center of each spaxel, which are reported in the {\tt SLITMAP} extension of the DRP output files. 

This last step relies on laboratory metrology measurements of the relative position between the IFU lenslets and the AG camera detectors, which were performed prior to commissioning. The focal plane assembly is designed so the relative position of these components remains stable \citep{herbst24}. On-sky verification using crowded fields shows that the systematic error in this metrology measurements is approximately $2.5\arcsec$, which corresponds to seven hundredths of a spaxel diameter \citep{haberle26}. The astrometric precession of the LVM data is therefore given by a combination of this systematic error in the focal plane metrology, and the error in the AG camera astrometry solutions.

\begin{figure*}[th]
\gridline{\fig{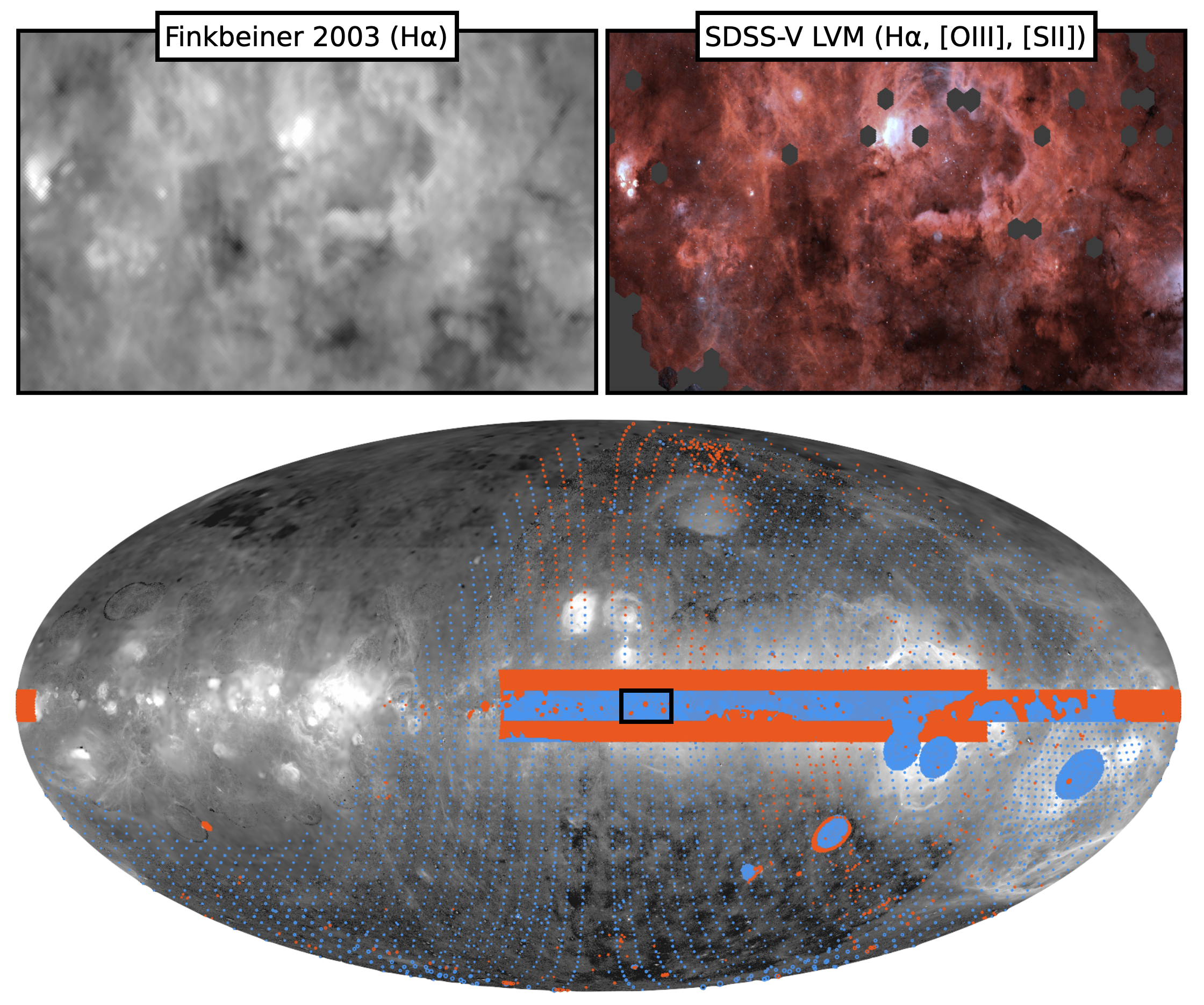}{0.99\textwidth}{}}
\vspace{-1.5\baselineskip}
\caption{ Survey progress as of October 20th 2026. The bottom panel shows the planned (red) and observed (blue) LVM tiles overlayed on the all-sky H$\alpha$ map of \cite{finkbeiner03}. The top panels show a $14^{\circ}\times 8^{\circ}$ zoom in region (outlined in black in the bottom panel), and compare the H$\alpha$ only \cite{finkbeiner03} map, which is mainly based on SHASSA \cite{gaustad01} survey data in this particular region, to a three color RGB composite of the SDSS-V LVM data build from the H$\alpha$, [OIII]$\lambda$5007, and [SII]$\lambda$6717 emission lines.
}
\vspace{0.5\baselineskip}
\label{fig:progress}
\end{figure*}

Figure \ref{fig:astrometry} presents the distribution of coordinate residuals for all GAIA stars detected in all coadded AG camera images over one complete night of LVM-I operations (MJD 61088). The histogram shows the residuals of each individual star, and the vertical lines show the average of the all residuals for each individual science exposure. The AG cameras' astrometric solutions show typical average absolute residuals of $\sim0.25\arcsec$, which are ten times smaller than the systematic error in the focal plane metrology measurements, and less than 1/100$^{th}$ of the spaxel diameter. 

Users of the data can therefore expect to observe systematic coordinate offsets at the 1/10$^{th}$ of a spaxel level \citep{haberle26}, and with a largely negligible (1/10$^{th}$ of a spaxel) random scatter around it (Figure \ref{fig:astrometry}). The direction of these offsets will depend on the IFU position angle used in a particular observation. Depending on the science application, these systematic offsets might be important to quantify (e.g. via comparison to other datasets), but for most science uses they might be tolerable as they are are well below the spatial resolution of the LVM data.

\subsection{Survey Progress}

The SDSS-V LVM program started survey operations in November 6th 2023. At the time of writing (October 20th 2026), the project has observed 20,428 science exposures over 15,548 tiles. This amounts to more than 36 million science spectra. The bottom panel of Figure \ref{fig:progress} shows the planned (red) and currently observed (blue) footprint of the LVM survey, overlaid on the all-sky H$\alpha$ map published by \cite{finkbeiner03}. It can be seen that Orion, the Gum nebula, and the bulk of the SMC and the LMC have been largely completed, and that extensive coverage of the main MW plane area is already available. Significant coverage of the LV galaxies and the all-sky sparse grid is also in hand.

The top two panels in Figure \ref{fig:progress} present a comparison between the SDSS-V LVM data and the \cite{finkbeiner03} H$\alpha$ map for a $14^{\circ} \times 9^{\circ}$ region in the Galactic midplane (black box in bottom panel). The \cite{finkbeiner03} map combines data from the WHAM survey ($1^{\circ}$ resolution and 0.15~R depth), the Virginia Tech Spectral line Survey \citep[VTSS,][1.6' resolution and 0.5~R depth]{dennison98}, and the Southern H$\alpha$ Sky Survey Atlas \citep[SHASSA][at 0.8' resolution and 2~R depth]{gaustad01} surveys. The H$\alpha$ map data in the comparison area comes from the latter. The SDSS-V LVM data shown for this area is color composite map built from the H$\alpha$, [OIII]$\lambda$5007, and [SII]$\lambda$6717 DAP measured line fluxes. This comparison highlights the transformative spectral richness and improved spatial resolution of the SDSS-V LVM data, over previous systematic ionized gas emission mapping efforts in the Milky Way.

%% file: 8-conclusions.tex
\section{Conclusions} \label{sec:conclusions}

In this article we have presented a technical overview of the SDSS-V Local Volume Mapper (LVM) project. We have outlined the design and characteristics of the LVM Instrument (LVM-I), the survey footprint and tiling strategy, the observing procedures, and the main components of the LVM data workflow, including the reduction and analysis frameworks, and the main data products that are produced by the project. We have also discussed the expected quality and stability of the data products based on commissioning and early science observations.

The LVM represents a unique capability for optical integral-field spectroscopy of the nearby universe. Its combination of ultra-wide field-of-view, broad wavelength coverage (3600–-9800~\AA), medium spectral resolution, and high physical spatial resolution enables the systematic mapping of the ionized interstellar medium and stellar populations across the Milky Way, the Magellanic Clouds, and nearby galaxies. In the Milky Way and Magellanic Clouds in particular, the survey probes physical scales associated with the direct interaction between stars and the surrounding ISM, providing access to the fundamental scales at which stellar feedback operates.

The custom-built LVM-I facility at Las Campanas Observatory was specifically designed to support this science case. The use of dedicated science, sky, and spectrophotometric telescopes enables simultaneous acquisition of calibration and science data over the large angular scales required by the survey. The static fiber system and thermally stabilized spectrographs provide excellent instrumental stability, minimizing calibration overheads and enabling robust long-term operations. The robotic observing framework, combined with automated scheduling, acquisition, guiding, and calibration procedures, allows the efficient mapping of the large survey footprint envisioned for SDSS-V which covers several thousands of square degrees.

The LVM data processing framework is designed to transform the raw observations into science-ready data products. The LVM pipelines produce calibrated spectra, emission-line measurements, stellar population parameters, and higher-level derived quantities for every observed spaxel in the survey. Together with visualization and archive tools distributed through the SDSS infrastructure, these products are intended to make the survey broadly accessible to the SDSS-V collaboration members, and to the whole astronomical community and the public through SDSS public data releases.

As survey operations continue, the LVM will produce an unprecedented spectroscopic view of the nearby universe across a wide range of environments and physical scales. The resulting dataset will provide a powerful resource for studies of the lifecycle of baryons in galaxies, including star formation, stellar feedback, gas ionization, chemical enrichment, and the coupling between stars and the interstellar medium. LVM officially began science operations in late 2023 and is expected to continue through at least 2027. Future SDSS-V data releases will progressively expand the publicly available dataset and associated analysis products, enabling increasingly comprehensive investigations of the Local Volume and the Milky Way ecosystem.
